\documentclass[aps,prd,nofootinbib,superscriptaddress,floatfix]{revtex4}
\usepackage{graphicx}
\usepackage{bm}
\usepackage{braket}
\usepackage{bbold}
\usepackage{amssymb}
\usepackage{amsmath,latexsym}
\usepackage{subfigure}
\usepackage{slashed}
\usepackage[T1]{fontenc}
\usepackage{multirow,array}
\usepackage{mathtools}
\usepackage{mathrsfs}
\usepackage{dsfont}
\usepackage[colorlinks=false,linktocpage=true]{hyperref}
\usepackage{hyperref}
\usepackage[utf8]{inputenc}
\usepackage[export]{adjustbox}
\usepackage{float}
\usepackage[makeroom]{cancel}
\usepackage{wrapfig}
\usepackage{ulem}
\usepackage{color}

\newcommand{\be}{\begin{equation}}
\newcommand{\ee}{\end{equation}}
\newcommand{\ba}{\begin{eqnarray}}
\newcommand{\ea}{\end{eqnarray}}
\newcommand{\lambdaS}{ s}
\newcommand{\sub}[1]{   \begin{subequations}
                        #1
                        \end{subequations} }

\newcommand{\di}{\!{\rm d}}
\newcommand{\la}{\langle}
\newcommand{\ra}{\rangle}

\newcommand{\fslash}[1] {{\not\! #1\,}}

\begin{document}

\newcommand*{\UConn}{Department of Physics, University of Connecticut,
    Storrs, CT 06269, U.S.A.}
\newcommand*{\BNL}{Department of Physics, 
    Brookhaven National Laboratory, Upton, NY 11973, U.S.A.}

\title{\boldmath
Chiral-odd GPDs in the bag model}
\author{Kemal Tezgin}\affiliation{\UConn}\affiliation{\BNL}
\author{Brean Maynard}\affiliation{\UConn}
\author{Peter Schweitzer}\affiliation{\UConn}
\date{April 2024}
\begin{abstract}
\noindent
A study of chiral-odd generalized parton distributions (GPDs) of the nucleon is 
presented in the bag model demonstrating that in this model all four chiral-odd 
GPDs are non-zero contrary to other claims in literature. The bag model results
for the GPDs $H_T^q(x,\xi,t)$, $E_T^q(x,\xi,t)$, $\tilde{H}_T^q(x,\xi,t)$ agree
with other models within a typical quark model accuracy. 
We present one of the few quark model calculations where polynomiality
is satisfied and the sum rule $\int dx\,\tilde{E}_T^q(x,\xi,t)=0$ holds. 
We confront our results with 
predictions from the large-$N_c$ limit, and with lattice QCD calculations. 
We conclude that the bag model successfully catches the main features of 
chiral-odd GPDs.
\end{abstract}
%
%
%

\maketitle

\section{Introduction}
\label{sec-1:Introduction}

The chiral-odd GPDs of the nucleon \cite{Diehl:2001pm} are difficult to access
experimentally as they cannot be studied in deeply virtual Compton scattering 
or hard-exclusive vector meson production \cite{Collins:1999un} which provide
the main source of information for chiral-even GPDs  \cite{Mueller:1998fv,Ji:1996ek,Radyushkin:1996nd,Collins:1996fb}.
In order to access chiral-odd GPDs experimentally it is necessary to consider
more complex final states like diffractive exclusive production of two vector 
mesons \cite{Ivanov:2002jj}, other reactions which are similarly challenging to 
measure \cite{Enberg:2006he,Pire:2009ap,ElBeiyad:2010pji,Pire:2015iza,
Boussarie:2016qop,Pire:2017lfj,Duplancic:2023kwe}, or approaches based on an 
effective handbag diagram dominance framework for pion electroproduction which 
are not rooted in collinear factorization \cite{Ahmad:2008hp,Goloskokov:2009ia,
Goloskokov:2011rd,Goloskokov:2013mba}. As a consequence,
the chiral-odd GPDs received less attention in literature than the chiral-even
GPDs, see Ref.~\cite{Ji:1998pc,Goeke:2001tz,Diehl:2003ny,Belitsky:2005qn,
Boffi:2007yc,Guidal:2013rya,Kumericki:2016ehc,dHose:2016mda} for reviews, 
and there are also fewer model studies of chiral-odd GPDs in literature.

In this work, we report a study of the chiral-odd GPDs in the bag model.
Historically, the bag model \cite{Chodos:1974je,Chodos:1974pn} has been 
often explored as a first theoretical approach for pioneering studies of
nucleon properties: from first studies of
structure functions \cite{Jaffe:1974nj}, to first studies of the transversity
parton distribution function (PDF) $h_1^q(x)$ \cite{Jaffe:1991kp}, 
to one of the earliest systematic model studies of transverse momentum dependent 
PDFs \cite{Avakian:2010br} as well as the first model study of chiral-even GPDs \cite{Ji:1997gm}. 
In view of this history, it is not surprising that to the best of our knowledge 
the bag model was also used for the first model study of chiral-odd GPDs in
Ref.~\cite{Scopetta:2005fg}.

In Ref.~\cite{Scopetta:2005fg} it was concluded that in the bag model three out of 
the four chiral-odd nucleon GPDs vanish and only one is non-zero, namely the one 
which reduces to $h_1^q(x)$ in the forward limit. Subsequent works
\cite{Pasquini:2005dk,Pasquini:2007xz,Lorce:2011dv,Chakrabarti:2015ama,
Maji:2017ill,Wakamatsu:2008ki,Kaur:2023lun,Liu:2024umn,Goldstein:2013gra,Dahiya:2007mt,Chakrabarti:2008mw}
revealed that in other models all four chiral-odd GPDs are non-zero. 
This gives rise to the question why the description of chiral-odd GPDs 
in the bag model would differ so distinctly from other models. 

It is natural to expect that different quark models give different results. 
But within their ``range of applicability'' one often observes that 
models agree with each other within a certain ``model accuracy.'' 
For instance in the case of transverse momentum dependent PDFs, the bag model 
is in line with other quark models within a model accuracy of about 20-40$\,\%$ 
which is inferred from comparisons of different models \cite{Boussarie:2023izj}.
But if one model \cite{Scopetta:2005fg} predicts three of the chiral-odd GPDs 
to be zero at variance with other models \cite{Pasquini:2005dk,Pasquini:2007xz,
Lorce:2011dv,Chakrabarti:2015ama,
Maji:2017ill,Wakamatsu:2008ki,Kaur:2023lun,Liu:2024umn,Goldstein:2013gra,Dahiya:2007mt,Chakrabarti:2008mw},
then this warrants a closer investigation. 
The goal of this work is to revisit the calculation of chiral-odd GPDs in the bag model.

This work is organized as follows.
In Sec.~\ref{Sec-2:GPDs} we will define the chiral-odd GPDs and briefly review
their properties. In Sec.~\ref{Sec-3:bag-model} we will introduce the bag model
and review its description of chiral-even GPDs following \cite{Ji:1997gm}. 
In Sec.~\ref{Sec-4:chiral-odd-GPDs-in-bag} we will derive the bag model expressions 
for the chiral-odd GPDs, and in Sec.~\ref{Sec-5:numerical-result} we will present 
the numerical results which show that all four chiral-odd GPDs are non-zero in 
the bag model. In Sec.~\ref{Sec-6:compare-to-models}, after explaining why the 
previous bag model study \cite{Scopetta:2005fg} missed three of the chiral-odd GPDs,
we compare our results to other models \cite{Pasquini:2005dk,Pasquini:2007xz,
Lorce:2011dv,Chakrabarti:2015ama,Maji:2017ill,Wakamatsu:2008ki,Kaur:2023lun,Liu:2024umn}
and show that they agree with other models within an accuracy similar to 
the quark-model description of transverse momentum dependent PDFs \cite{Boussarie:2023izj}. 
In Sec.~\ref{Sec-7:compare-to-large-Nc} we will confront the bag model results to 
predictions from large-$N_c$ limit in QCD 
\cite{Schweitzer:2016jmd,Schweitzer:2015zxa}, and Sec.~\ref{Sec-8:compare-to-lattice}
is dedicated to a comparison of our results with lattice QCD calculations of 
chiral-odd GPDs based on the quasi distribution method \cite{Alexandrou:2021bbo}.
The Sec.~\ref{Sec-9:FFs} discusses the tensor form factors, and
Sec.~\ref{Sec-10:conclusions} presents the conclusions. 
Technical details of the model calculation can be found in the Appendices.

\section{Definition and properties of chiral-odd GPD\lowercase{s}}
\label{Sec-2:GPDs}

Let $p,\,p'$ and $\lambdaS,\lambdaS'$ denote respectively 
the incoming, outgoing nucleon momenta and polarization vectors.
We introduce the four-vectors $P=\frac{1}{2}(p+p')$ and $\Delta=p^\prime-p$,
and choose the spatial component of the lightcone direction along $z$-axis and 
define lightcone coordinates as $z^\pm = \frac{1}{\sqrt{2}}(z^0\pm z^3)$ and 
$\vec{z}_T=(z^1,z^2)$. 

In this notation, quark GPDs are generically defined through the matrix of non-local 
quark operators of the type \cite{Mueller:1998fv, Ji:1996ek, Radyushkin:1996nd}
\be\label{Eq:abbreviation}
    {\cal M}_{\lambdaS^\prime\lambdaS}[\Gamma] =
    P^{+}\int\frac{dz^{-}\!\!}{2\pi}\;e^{ixP^{+}z^{-}}
    \bra{p',\lambdaS'}\bar\psi_q(-\frac{z}{2})
    \mathcal{W}(-\frac{z}{2},\frac{z}{2})\,\Gamma
    \psi_q(\frac{z}{2})\ket{p,\lambdaS}\bigg|_{z^{+}=0,\vec{z}_T=0}
\ee
where the Wilson line is given by 
$\mathcal{W}(a,b)=\mathcal{P}\,\exp(ig\int_b^a dx_\mu\,A^\mu)$ with 
$\mathcal{P}$ denoting the path-ordering operator. In the cases
$\Gamma=\gamma^+$ or $\gamma^+\gamma_5$, one deals with the chiral-even twist-2 GPDs 
$H^q$, $E^q$ or $\tilde{H}^q$, $\tilde{E}^q$ which are functions of the variables 
$x$, $\xi$, $t$ with $\xi = (p^{+}-p'^{+})/(p^{+}+p'^{+})$ and $t=\Delta^2$.
In this work, we are specifically interested in the case $\Gamma=i\sigma^{+j}$ with 
the transverse indices $j=1,\,2$ which corresponds to chiral-odd twist-2 GPDs defined 
as \cite{Diehl:2001pm}
\ba
     \label{oddGPDs}
    {\cal M}_{\lambdaS^\prime\lambdaS}[\,i\sigma^{+i}\,] &=& \bar u(p',\lambdaS')
    \bigg[H_T^q\; i\sigma^{+i}+\tilde H_T^q\;\frac{P^{+}\Delta^i - 
    \Delta^{+}P^i}{m^2}+E_T^q\;\frac{\gamma^{+}\Delta^i-\Delta^{+}\gamma^i}{2m}
    +\tilde E_T^q\;\frac{\gamma^+ P^i- P^+\gamma^i}{m}\bigg]u(p,\lambdaS)\, ,
\ea
where nucleon spinors are normalized as $\bar{u}(p,s')u(p,s)=2\,m\,\delta_{ss'}$
and $m$ denotes the nucleon mass, and we do not indicate the variables $x,\,\xi,\,t$. 
Notice that GPDs depend on the renormalization scale which we suppress for brevity.

Time reversal invariance implies a specific behavior of the GPDs 
under the transformation $\xi\rightarrow -\xi$ according to
\begin{align}
    F^q(x,\xi,t) &= \phantom{-} F^q(x,-\xi,t) \;\;\text{ for  } \;\;
    F^q= 
    H_T^q,\tilde H_T^q, E_T^q\, ,
    \label{skewness1} \\
    F^q(x,\xi,t) &= -F^q(x,-\xi,t) \;\;\text{ for  } \;\;
    F^q=\tilde E_T^q\, .\label{skewness2}
\end{align}
In the forward limit $p'\to p$ the variables $\xi$ and $t$ vanish, and  
$H^q_T(x,0,0) = h_1^q(x)$ becomes the transversity PDF. The GPDs 
$E^q_T(x,0,0)$ and $\tilde{H}^q_T(x,0,0)$ have well-defined forward
limits but drop out in Eq.~(\ref{oddGPDs}) in the forward limit and as
a consequence do not correspond to PDFs, while $\tilde{E}^q_T(x,0,0)=0$ 
due to Eq.~(\ref{skewness2}). 

The Mellin moments of GPDs $\int dx\,x^{N-1}F^q(x,\xi,t)$ are even polynomials 
in $\xi$ for all twist-2 GPDs except $\tilde{E}_T^q(x,\xi,t)$ in which case they are
odd. The first Mellin moments correspond to the tensor current form factors 
according to
\ba
\label{odd-first-moment}
  \int dx\,H_T^q(x,\xi,t) = H_T^q(t), \quad 
  \int dx\,E_T^q(x,\xi,t) = E_T^q(t), \quad
  \int dx\,\tilde{H}_T^q(x,\xi,t) = \tilde{H}_T^q(t), \quad
  \int dx\,\tilde{E}_T^q(x,\xi,t) = 0\,.
\ea
Notice that the lowest Mellin moment of $\tilde{E}_T^q(x,\xi,t)$ vanishes and
there is no form factor associated with this GPD in Eq.~(\ref{odd-first-moment})
which follows directly from Eq.~(\ref{skewness2}). 
For simplicity, we use the same notation for GPDs and form factors as both can 
unambiguously be distinguished due to the different numbers of arguments.  

A set of strong, intermediate, and weak inequalities for chiral-odd GPDs 
has been derived for $x>\xi$ in Ref.~\cite{Kirch:2005in}. 
These inequalities are a generalization of the Soffer bound \cite{Soffer:1994ww}
and formulated in terms of the unpolarized, helicity, and transversity 
PDFs $f_1^q(x)$, $g_1^q(x)$, $h_1^q(x)$. The strong inequalities are
given by \cite{Kirch:2005in}
\ba\label{ineq:HT}
     |H_T^q(x,\xi,t)| & \le & \; \frac{\alpha}{4} \;\sum\limits_{k=1}^3\,
     \Biggl[    
     \biggl(R^q_k(x_+)R^q_k(x_-)+R^q_2(x_+)R^q_2(x_-)\biggr)^2+\nu^2\xi^2
     \biggl(R^q_k(x_+)R^q_2(x_-)+R^q_2(x_+)R^q_k(x_-)\biggr)^2\,\Biggr]^{1/2} , \;\;
     \\
\label{ineq:tildeET}
     |\tilde{E}_T^q(x,\xi,t)| & \le & \frac{\nu}{4\alpha}\;\sum\limits_{k=1}^3
     \Biggl[    
     \biggl(
     \frac{R^q_k(x_+)R^q_2(x_-)}{1+\xi}+
     \frac{R^q_2(x_+)R^q_k(x_-)}{1-\xi}\biggr)^2+4\nu^2\alpha^4\xi^2
     \biggl(R^q_2(x_+)R^q_2(x_-)\biggr)^2\,\Biggr]^{1/2} , 
     \\
\label{ineq:ET}
     |E_T^q(x,\xi,t)| & \le & \frac{\nu}{4\alpha}\;\sum\limits_{k=1}^3
     \Biggl[    
     \biggl(
     \frac{R^q_k(x_+)R^q_2(x_-)}{1+\xi}+
     \frac{R^q_2(x_+)R^q_k(x_-)}{1-\xi}\biggr)^2+4\nu^2\alpha^4
     \biggl(R^q_2(x_+)R^q_2(x_-)\biggr)^2\,\Biggr]^{1/2} ,
     \\
\label{ineq:ildeHT}
     |\tilde{H}_T^q(x,\xi,t)| & \le & \frac{\nu^2}{2\alpha}\:
     \Bigl(R^q_2(x_+)R^q_2(x_-)\Bigr)^2 ,
\ea
where
\ba
&&   R^q_1(x) = \sqrt{f_1^q(x)+g_1^q(x)+2h_1^q(x)}, \quad
     R^q_2(x) = \sqrt{f_1^q(x)-g_1^q(x)}, \quad
     R^q_3(x) = \sqrt{f_1^q(x)+g_1^q(x)-2h_1^q(x)}, \nonumber\\
&&   x_+ = \frac{x+\xi}{1+\xi} , \quad
     x_-= \frac{x-\xi}{1-\xi} , \quad
     \alpha = \frac{1}{\sqrt{1-\xi^2}}, \quad
     \nu =  \frac{2m}{\alpha(t_0-t)},\ \quad
     t_0 = -\,\frac{4m^2\xi^2}{1-\xi^2}\,.
\ea
The intermediate and weak inequalities are less stringent, and
will not be discussed in this work.

In the limit $\xi\to0$, the GPDs have probabilistic interpretations:
their Fourier transforms describe spatial distributions of partons 
(carrying a fraction $x$ of nucleon momentum) in the plane transverse 
to the nucleon momentum \cite{Burkardt:2002hr,Burkardt:2005hp}.
For more details on GPDs, see the reviews 
\cite{Ji:1998pc,Goeke:2001tz,Diehl:2003ny,Belitsky:2005qn}.

\section{Bag model and description of GPD\lowercase{s}}
\label{Sec-3:bag-model}

In the bag model, the nucleon is described by placing $N_c=3$ free quarks 
in a color singlet state inside a spherical~cavity (``bag'') of radius $R$.
The quarks are confined by appropriately imposed boundary conditions
\cite{Chodos:1974je,Chodos:1974pn}. In the bag rest frame, the 
coordinate-space and momentum-space ground-state single-quark 
wave functions are given by
\ba\label{Eq:bag-wave-function}
    \psi_{s}(t,\vec{r}) &=& e^{-i\varepsilon_0t}
    \,\phi_s(\vec{r})
        \, , \;\;\;
    \phi_s(\vec{r})   = \frac{A}{\sqrt{4\pi}}
    \left(\begin{array}{l}
        j_0(\omega_0r/R)\,\chi_s \phantom{\displaystyle\frac11}\\
        j_1(\omega_0r/R)\,i\vec{\sigma}\cdot\vec{e}_r\chi_s
    \end{array}\right) \, , \\
    \label{wave-function-momentum}
    \psi_{s}(t,\vec{k}) &=& e^{-i\varepsilon_0t}\,\phi_s(\vec{k})
    \, , \;\;\;
    \phi_s(\vec{k})   =  \sqrt{4\pi}\:A\:R^3\;
    \left(\begin{array}{l}
        t_0(k)\,\chi_s \phantom{\displaystyle\frac11}\\
        t_1(k)\,\vec{\sigma}\cdot\vec{e}_k\chi_s
    \end{array}\right) \, , 
\ea
where the single-quark energy is given by $\varepsilon_0=\omega_0/R$
and $\omega_0\approx 2.04$ is the lowest positive solution 
of the transcendental equation $\omega = (1-\omega)\,\tan\omega$. In these
expressions we assume the quarks to be massless, and $\sigma^i$ are $2\times2$
Pauli matrices with $\chi_s$ denoting the two-component Pauli spinors.
The constant $A = \omega_0/\sqrt{2R^3 j_0(\omega_0)^2\omega_0(\omega_0-1)}$ 
ensures the normalization 
$\int\di^3x\,\phi^\dag_{s^\prime}(\vec{x}^{\,})\,\phi^{ }_{s^{ }}(\vec{x}^{\,})
=\delta_{s^\prime\!s}$. The coordinate-space
functions $j_l(\omega_0r/R)$ for $l=0,1$ are spherical Bessel functions.
The momentum-space functions $t_l(k)$ for $l=0,1$ are given by
\begin{equation}
    t_l(k)=\int_{0}^{1} du\,u^2\,j_l(ukR)\,j_l(u\omega_i) \, .
\end{equation}
The bag radius is fixed by the physical
value of the nucleon mass $m$ via the relation $m=4N_c \omega_0/(3R)$.

SU(4) spin-flavor symmetry relates nucleon matrix elements of quark operators 
to their single quark counterparts through spin-flavor factors: 
$N_q$ for spin-independent and $P_q$ for spin-dependent nucleon
matrix elements. For the proton 
$N_u = \frac{N_c+1}{2}$, 
$N_d = \frac{\;N_c-1}{2}$, 
$P_u = \frac{N_c+5}{6}$, 
$P_d = \frac{-N_c+1}{6}$.
For the neutron, the labels $u$ and $d$ are interchanged~\cite{Karl:1984cz}.

The first model study of quark GPDs was presented in the bag model
\cite{Ji:1997gm} where chiral-even GPDs were studied. 
We shall briefly review the method of Ref.~\cite{Ji:1997gm}.
For the calculation, it is convenient to choose the Breit frame where the 
incoming and outgoing nucleon four-momenta, skewness and $t$ are given by
\begin{equation}
    \label{Eq:Breit-frame}
    p^\mu = (\bar{m},-\vec{\Delta}/2)\, , \qquad 
    p^{\prime\mu} = (\bar{m},\vec{\Delta}/2)\, , \qquad
    \Delta^\mu = (0,\vec{\Delta}) = (0,\Delta_x,\Delta_y,\Delta_z), \quad
    \xi=-\frac{\Delta_z}{2\bar{m}}\, , \qquad
    t=-\vec{\Delta}^2
\end{equation}
with $\bar{m}^2=P^2=m^2-t/4$. Following \cite{Ji:1997gm}, the 
coordinate-space wave function (\ref{Eq:bag-wave-function}) is transformed 
from the rest frame to a frame where the nucleon moves with velocity 
$\vec{v}$ by means of a Lorentz boost \cite{Betz:1983dy} as follows 
\be\label{Eq:boost}
    \psi_{\vec{v}}(t,\,\vec{r})
    =
    S(\Lambda_{\vec{v}}) \,
    \psi\bigl((t-\vec{v}\cdot\vec{r})\cosh w, \,
    \vec{r}+\hat{v}\cdot\vec{r}\,(\cosh w-1)\hat{v}-\vec{v}t\cosh w\bigr)\,,
\ee
where $w=\tanh^{-1}|\vec{v}|$ and
$S(\Lambda_{\vec{v}})=\cosh\frac{w}{2}+(\hat{v}\cdot\vec{\alpha})\sinh\frac{w}{2}$
with $\alpha^i=\gamma^0 \gamma^i$ and the unit vector 
$\hat{v}=\vec{v}/|\vec{v}|$. The velocity of the incoming nucleon is 
$\vec{v}=-\vec{\Delta}/2\bar{m}$ such that 
\begin{equation}
     \cosh w=\frac{\bar{m}}{m},\quad \mbox{and} \quad\sinh w=\frac{|\vec{\Delta}|}{2m}.
\end{equation}
The transformed momentum-space wave function is obtained from the Fourier 
transformation \cite{Ji:1997gm}
\begin{equation}
     \psi_{\vec{v}}(t,\vec{r})=S(\Lambda_{\vec{v}})\int\frac{d^3\vec k}{(2\pi)^3}\,
     \mbox{exp}(-i\tilde{k}^\mu x_\mu)\;\varphi(\vec{k})\, , \quad
     \tilde{k}^\mu = (\tilde{\epsilon}_0, \vec{\tilde k})\, , \quad
     \vec{\tilde k} = \vec{\tilde{k}}_\parallel + \vec{\tilde{k}}_\perp
\end{equation}
where $x^\mu=(t,\vec{x})$ and 
$\tilde{\epsilon}_0=(\epsilon_0+\vec{k}\cdot\vec{v})\cosh w$,
$\vec{\tilde{k}}_\perp=\vec{k}_\perp$, 
$\vec{\tilde{k}}_\parallel=(\epsilon_0\vec{v}+\vec{k}_\parallel)\cosh w$ 
where $\vec{k}_\perp\cdot\hat{v}=0$ 
and $\vec{k}_\parallel=(\vec{k}\cdot\hat{v})\hat{v}$ 
are respectively perpendicular and parallel with respect to 
the nucleon velocity.

With the relativistic motion of the nucleon taken into account,
the bag model expressions for the matrix elements in
Eq.~(\ref{Eq:abbreviation}) are given by \cite{Ji:1997gm}
\begin{equation}\label{GPDmatrixelements}
   {\cal M}_{\lambdaS^\prime\lambdaS}[\Gamma] =
   \frac{2\bar{m}}{\cosh w}\;
   \frac{\bar{m}}{1-(\cosh w - 1)\Delta_z^2/t}
   \int\frac{d^3\vec k}{(2\pi)^3}\delta(k_z-K)
   {\cal A}_{\lambdaS^\prime\lambdaS}[\Gamma] \, .
\end{equation} 
The expressions  $K$ in the 
argument of the $\delta$-function and ${\cal A}_{\lambdaS^\prime\lambdaS}[\Gamma]$
in Eq.~(\ref{GPDmatrixelements}) are defined as
\ba\label{GPDmatrixelements2}
   {\cal A}_{\lambdaS^\prime\lambdaS}[\Gamma] &=&
   \Big\{\varphi^\dagger_{s'}(k')S(\Lambda_{-\vec{v}})\gamma^0\Gamma 
   S(\Lambda_{\vec{v}})\varphi_s(k)\Big\} \\
   \label{GPDmatrixelements3}
    K &=& \frac{\bar{m}}{1-(\cosh w-1)\Delta_z^2/t}\Big[x-\frac{1}{2\bar{m}}
    \Big( (2\frac{w_0}{R}+\tilde\Delta_z)\cosh w + |\vec{\tilde\Delta}|\sinh w 
    -\frac{2 \vec{\Delta}_\perp\cdot\vec{k}_\perp}{t}\Delta_z(\cosh w -1)\Big)
    \Big]\bigg)\, ,
\ea
where $k^\prime = |\vec{k}^\prime|$ with
$\vec{k}^\prime = \vec{k} + \vec{\tilde\Delta}$. The term $\vec{\tilde\Delta}$ denotes the effective 
momentum transfer to the active quark which needs to be distinguished from
the momentum transfer to the entire nucleon and is defined as \cite{Ji:1997gm}
\begin{equation}\label{Eq:tildeDelta}
    \vec{\tilde\Delta}=\eta\frac{\vec{\Delta}}{\cosh w} \, ,
\end{equation}
%
%
%
%
where $\eta$ should take the value 
$\eta_0=1-\epsilon_0/M$. In this work we follow 
Ref.~\cite{Ji:1997gm} where it was proposed to treat $\eta$ as 
a free model parameter to remedy a deficiency,  namely 
the fact that in the above model treatment the momentum 
transfer $\vec{\Delta}$ to the {\it nucleon} is absorbed
entirely by one single (``active'') quark while 
the other (``spectator'') quarks receive no momentum
transfer. In quark models, one would expect each of 
the 3 quarks to receive $\frac13$ of the momentum transfer. 
In nature, correlations between the quarks ensure equal
sharing of the momentum transfer among the constituents. 
Such correlations are not included in independent particle models like the bag model. One way to approximately
address this problem is to allow smaller values $\eta < \eta_0$.
To that end in \cite{Ji:1997gm} it was proposed to fix $\eta$
from fits to the $t$-dependence of electromagnetic factors.
%
%
%
Using the above-described method, the chiral even GPDs 
$H^q(x,\xi,t)$, $E^q(x,\xi,t)$, $\tilde{H}^q(x,\xi,t)$, $\tilde{E}^q(x,\xi,t)$ 
were computed in the bag model in Ref.~\cite{Ji:1997gm}, and independently 
confirmed in Ref.~\cite{Tezgin:2020vhb}.

\section{Chiral-odd GPD\lowercase{s} in the bag model}
\label{Sec-4:chiral-odd-GPDs-in-bag}

In this section, we present our results for chiral-odd GPDs in 
the bag model. In order to determine the four chiral-odd GPDs 
in Eq.~(\ref{oddGPDs}) we need to establish and solve a set of
four linearly independent equations.
In the following we outline the main steps, and show the detailed
(often lengthy) expressions in the Appendices.

In order to derive the four linearly independent equations, 
we first evaluate in the bag model the nucleon matrix elements 
${\cal M}_{\lambdaS^\prime\lambdaS}[i\sigma^{+j}]$ on the
left-hand side of Eq.~(\ref{oddGPDs}) for $j=1,\,2$, cf.\
Eq.~(\ref{GPDmatrixelements}).
The resulting model expressions are $2\times2$ matrices in nucleon spinor 
indices, see App.~\ref{App:model-expressions-for-matrix-elements}.
Then we take traces of these expressions with different Pauli matrices
$\sigma^k$ with $k=0,\,1,\,2,\,3$ where $\sigma^0$ stands for the unit 
$2\times2$ matrix. This corresponds to projections on various nucleon spin 
sectors, namely unpolarized or polarized along the $x$, $y$, $z$-axis. 
The four projections give results for four bag-model quantities 
$T_i^q(\vec{\Delta})$ given in App.~\ref{App:Tq}.
Finally, we evaluate the nucleon spinor structures on the right-hand side of 
Eq.~(\ref{oddGPDs}) and take analogous projections in the nucleon spin sector,
see App.~\ref{App:nucleon-spinor-expressions} for details.

In this way we obtain the system of four linear equations
\begin{alignat}{6}
    \label{T1} T_1^q(\vec{\Delta}) = &&
      \frac{\Delta_x}{2m}\;H_T^q(x,\xi,t) &&
    + \frac{\Delta_x}{2m} \; E_T^q(x,\xi,t) &&
    + \frac{\bar m^2\Delta_x}{m^3}\;\tilde H_T^q(x,\xi,t) 
    \, \\
    \label{T2} T_2^q(\vec{\Delta}) = &&
      \Big[1 + \frac{\Delta_y^2}{4m(\bar m+m)}\Big]\;H_T^q(x,\xi,t) &&
    - \frac{\Delta_x^2+\Delta_z^2}{4m^2}\;E_T^q(x,\xi,t) &&
    - \frac{\bar m\Delta_z}{2m^2}\;\tilde E_T^q(x,\xi,t)
    \, , \\
    \label{T3} T_3^q(\vec{\Delta}) = &&
    - \Big[1 + \frac{\Delta_x^2}{4m(\bar m+m)}\Big]\;H_T^q(x,\xi,t) && 
    + \frac{\Delta_y^2+\Delta_z^2}{4m^2}\;E_T^q(x,\xi,t) &&
    + \frac{\bar m\Delta_z}{2m^2} \; \tilde E_T^q(x,\xi,t)
    \, ,\\
    \label{T4} T_4^q(\vec{\Delta}) = &&
    - \frac{\Delta_x \Delta_z}{4m(\bar m+m)} \; H_T^q(x,\xi,t) &&
    - \frac{\Delta_x \Delta_z}{4m^2} \; E_T^q(x,\xi,t) &&
    - \frac{\bar m \Delta_x}{2m^2} \; \tilde E_T^q(x,\xi,t)
    \, ,
\end{alignat}
with the explicit model expressions for the terms $T_i^q(\vec{\Delta})$
listed in Appendix~\ref{App:Tq}.
The individual chiral-odd GPDs are obtained in terms of linear 
combinations of the $T_i^q(\vec{\Delta})$ for $i=1,\,2,\,3,\,4$
by solving the linear system of Eqs.~(\ref{T1}--\ref{T4}).

Under the transformation $\vec{\Delta} \to -\,\vec{\Delta}$,
the $T_1^q(-\vec{\Delta})=-T_1^q(\vec{\Delta})$ is odd, while 
$T_i^q(-\vec{\Delta})=T_i^q(\vec{\Delta})$ for $i=2,\,3,\,4$
are even as can be seen in App.~\ref{App:Tq}. 
From these properties of the $T_i^q(\vec{\Delta})$, it follows 
that the chiral-odd GPDs $H_T^q,\;E_T^q,\,\tilde H_T^q$ obtained 
from Eqs.~(\ref{T1}--\ref{T4}) are even functions of $\xi$ while 
$\tilde E_T^q$ is odd, i.e.\ the bag model complies with 
the Eqs.~(\ref{skewness1},~\ref{skewness2}).
It is not immediately apparent from the model expressions,
but the obtained GPDs do not depend on the components of the
momentum transfer $\Delta_x$ or $\Delta_y$,
but only on the modulus of $\vec{\Delta}_\perp = (\Delta_x,\,\Delta_y)$.
With the third component $\Delta_z$ being related to $\xi$ in the 
Breit frame by Eq.~(\ref{Eq:Breit-frame}), this implies that 
for a fixed value of $\xi$, the GPDs are manifestly functions 
of $t$ which in the Breit frame is given by 
$t = -\vec{\Delta}{ }^2$, see Eq.~(\ref{Eq:Breit-frame}).

Before presenting our results, it is important to comment on the
scale. In QCD, GPDs depend on the renormalization scale, and results 
from models also have an ``effective scale''. The results from the bag 
model and many quark models refer to scales below the nucleon mass
and possibly as low as $\mu_0 = {\cal O}(500\,{\rm MeV})$
\cite{Stratmann:1993aw,Boffi:2009sh,Pasquini:2011tk,Pasquini:2014ppa}.

\section{Numerical results}
\label{Sec-5:numerical-result}

In this section we discuss the bag model results for chiral-odd GPDs.
Generally, quark models are applicable in the valence-$x$ region. We therefore choose 
to display throughout the $x$-region $0 < x < 0.8$ keeping in mind that the model 
results can be expected to be most reliable for $0.1\lesssim x \lesssim 0.7$. 
It is important to remember that the bag model gives rise to unphysical contributions for $x>1$ 
which are numerically very small and a consequence of the fact that the initial and final 
bag model nucleon states are not good momentum 
eigenstates~\cite{Ji:1997gm}.\footnote{It is worth recalling that the bag gives also 
    rise to unphysical (antiquark) GPDs in the region of $x<0$ which must be taken 
    into account when evaluating Mellin moments, verifying polynomiality properties, 
    and proving sum rules in the bag model \cite{Ji:1997gm}.
    \label{footnote-unphysical-qbar}} 

\begin{figure}[t!]
\begin{centering}
\includegraphics[width=4.2cm]{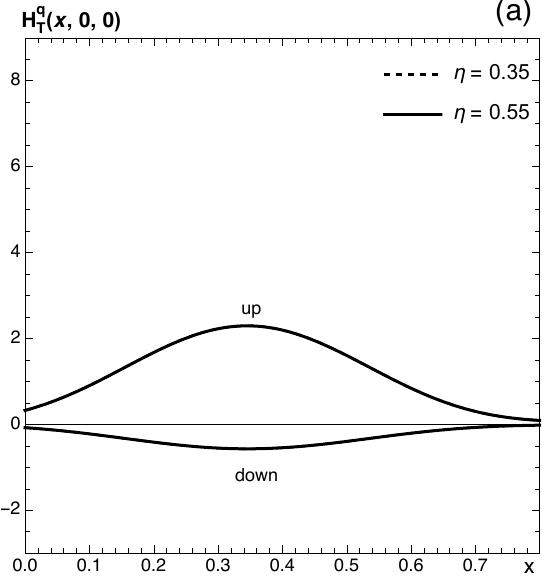} \
\includegraphics[width=4.2cm]{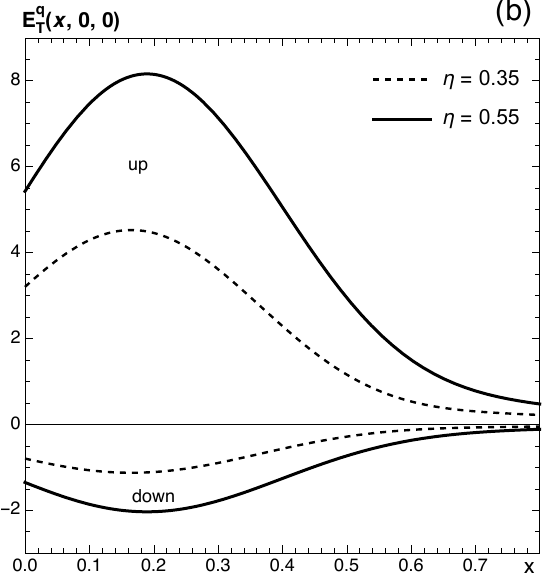} \
\includegraphics[width=4.2cm]{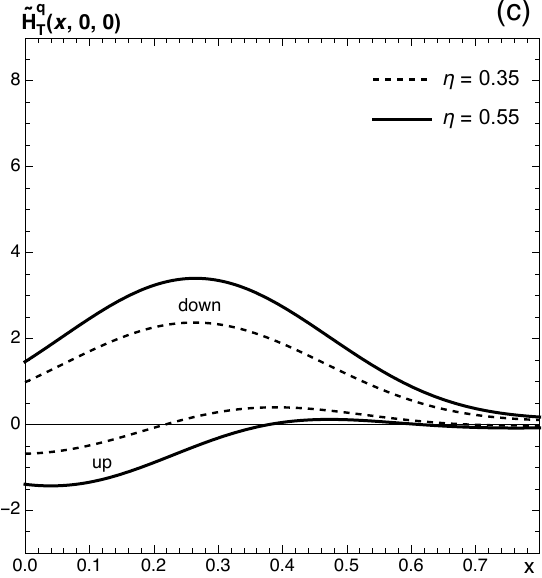} \
\includegraphics[width=4.2cm]{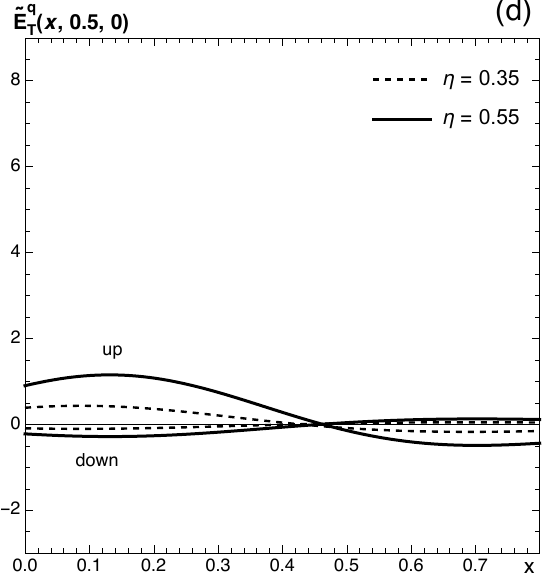} 
\par\end{centering}
\caption{\label{Fig1:ChiralOddForward}
The chiral-odd GPDs 
(a) $H_T^q(x,0,0)$,
(b) $E_T^q(x,0,0)$,
(c) $\tilde{H}^q_T(x,0,0)$, 
(d) $\tilde{E}^q_T(x,0.5,0)$ in bag model 
obtained in this work for the parameters $\eta=0.35$
and $0.55$ with the latter to be adopted for
the following figures. All GPDs are shown in forward
limit $\xi=0$ and $t=0$ except $\tilde{E}_T(x,0.5,0)$ 
is shown at $t=0$ analytically continued to $\xi=0.5$ 
as it vanishes at $\xi=0$.}
\end{figure}

In Figs.~\ref{Fig1:ChiralOddForward}a-c we show our results 
in the forward limit $t=0$ and $\xi=0$ 
for $H^q_T(x,\xi,t)$, $E^q_T(x,\xi,t)$, $\tilde{H}^q_T(x,\xi,t)$. 
In order to visualize the relative sizes of the GPDs, we choose 
the same scale in each panel in Fig.~\ref{Fig1:ChiralOddForward}.
Following Ref.~\cite{Ji:1997gm} we plot both choices for the 
parameter $\eta=0.35$ and $\eta=0.55$. We recall that $\eta=0.35$
yields a better description of electromagnetic form factors at 
larger momentum transfers $(-t)\gtrsim 1\,{\rm GeV}^2$ while
$\eta=0.55$ yields a better description in the region of
smaller $(-t) \lesssim 0.5\,{\rm GeV}^2$ \cite{Ji:1997gm}.

In Fig.~\ref{Fig1:ChiralOddForward}a we see that $H_T^q(x,0,0)$ is 
sizable and practically independent of the parameter $\eta$.
In fact, the curves for $\eta=0.35$ or $\eta=0.55$ are practically 
indistinguishable in Fig.~\ref{Fig1:ChiralOddForward}a.
The flavor dependence of $H_T^q(x,0,0)$ is given by the
ratio $u:d = 4:(-1)$ due to the spin-flavor factors $P_q$.
In the forward limit $H_T^q(x,0,0)=h_1^q(x)$ 
coincides with transversity computed first in the bag model 
in \cite{Jaffe:1991kp}.

Even though they have no PDF counter parts, the GPDs $E_T(x,0,0)$ and 
$\tilde{H}^q_T(x,0,0)$ have nevertheless well-defined forward limits as displayed
in Figs.~\ref{Fig1:ChiralOddForward}b and \ref{Fig1:ChiralOddForward}c.
Both GPDs show sensitivity to the numerical value of the parameter $\eta$ within 
about $30\,\%$ which is a typical ``theoretical accuracy'' 
of quark models in their range of applicability, i.e., in the valence-$x$ 
region $0.1\lesssim x \lesssim 0.7$
\cite{Stratmann:1993aw,Boffi:2009sh,Pasquini:2011tk,Pasquini:2014ppa,Bastami:2020asv}.
The sensitivity on the parameter $\eta$ observed in
Figs.~\ref{Fig1:ChiralOddForward}b and \ref{Fig1:ChiralOddForward}c
can hence be viewed as being within the model dependence 
encountered in quark models. 
$E_T^q(x,0,0)$ is the largest of all chiral-odd GPDs, while
$\tilde{H}_T^q(x,0,0)$ is of comparable magnitude to $H_T^q(x,0,0)$.
The flavor dependence of $E_T^q(x,0,0)$ is qualitatively similar 
to that of $H_T^q(x,0,0)$ and follows the $u:d = 4:(-1)$ ratio. The flavor dependence of 
$\tilde{H}_T^q(x,0,0)$ in Fig.~\ref{Fig1:ChiralOddForward}c is distinct 
from that of the other chiral-odd GPDs in that only in this case is
the $d$-flavor larger than the $u$-flavor, whereby the latter 
exhibits a node around $x\approx 0.2\mbox{-}0.3$ depending on the
choice for $\eta$, see Fig.~\ref{Fig1:ChiralOddForward}c.

The GPD $\tilde{E}_T^q(x,\xi,t)$ is a special case. 
It vanishes for $\xi=0$ due to the property (\ref{skewness2})
for any value of $t$.
In Fig.~\ref{Fig1:ChiralOddForward}d we therefore cannot show 
$\tilde{E}_T^q(x,\xi,t)$ for $t=0$ and $\xi=0$ as for the other GPDs.
Instead we depict $\tilde{E}_T^q(x,\xi,t)$ for $\xi=0.5$ continued 
analytically to $t=0$ which is achieved by choosing appropriate 
imaginary values for the components $\Delta_\perp^i$. Any GPD can 
be analytically continued in this way. 
The expressions at $t=0$ and $\xi\neq0$ are still
entirely real and develop no imaginary parts. Further 
applications of this analytic continuation technique 
for GPDs in another model can be found in
\cite{Schweitzer:2002yu,Schweitzer:2002nm,Schweitzer:2003ms,Ossmann:2004bp}.
Despite choosing the sizable value for $\xi=0.5$ (for
smaller $\xi$ this GPD decreases until it becomes zero
for $\xi=0$), the GPD $\tilde{E}_T^q(x,\xi,t)$ is by far 
the smallest of the chiral-odd GPDs. For this GPD, the
$u$- and $d$-flavors have opposite signs and exhibit nodes 
around $x\approx 0.4\mbox{-}0.5$. The nodes are expected 
features because this GPD must integrate to zero. The 
positive and negative contributions in
Fig.~\ref{Fig1:ChiralOddForward}d largely cancel each 
other but it should be kept in mind that the small
but non-negligible unphysical contributions in the
regions outside $0<x<1$ must be included in 
$\int dx\,\tilde{E}_T^q(x,\xi,t)$ for an
exact cancellation, cf.\ footnote~\ref{footnote-unphysical-qbar}.

It cannot be expected from quark models to describe GPDs or form factors 
for large $(-t)>1\,{\rm GeV}^2$ because they refer to scales lower than that
\cite{Stratmann:1993aw,Boffi:2009sh,Pasquini:2011tk,Pasquini:2014ppa}.
In our study we will therefore restrict ourselves to values of 
$(-t)<1\,\rm GeV^2$ and set from now on the parameter $\eta=0.55$, which was 
optimized to provide a better description of electromagnetic form factors at 
smaller $(-t)\lesssim0.5\,\rm GeV^2$ 
(whereas the choice $\eta=0.35$ was optimized to yield a better description 
of electromagnetic form factors for $(-t)\gtrsim 1\,{\rm GeV}^2$)
\cite{Ji:1997gm}.

\begin{figure}[t!]
\begin{centering}
\includegraphics[width=4.2cm]{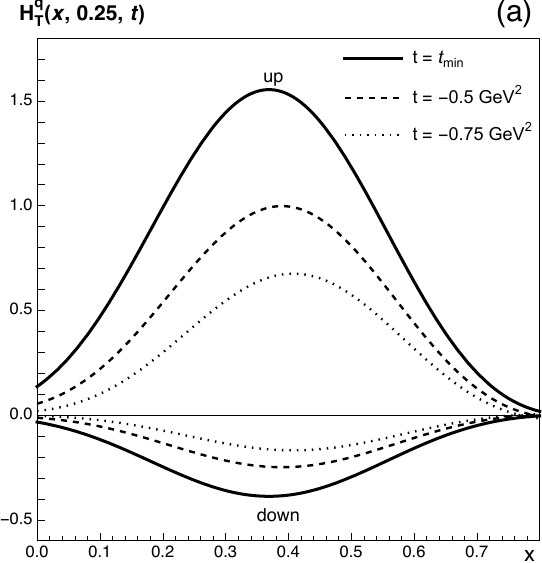} \
\includegraphics[width=4.2cm]{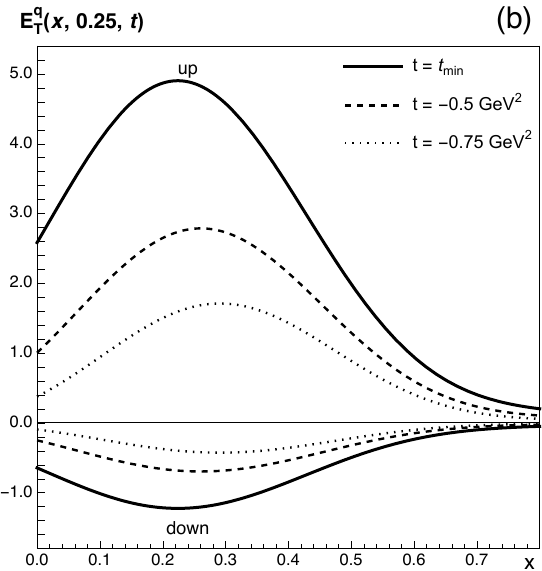} \
\includegraphics[width=4.2cm]{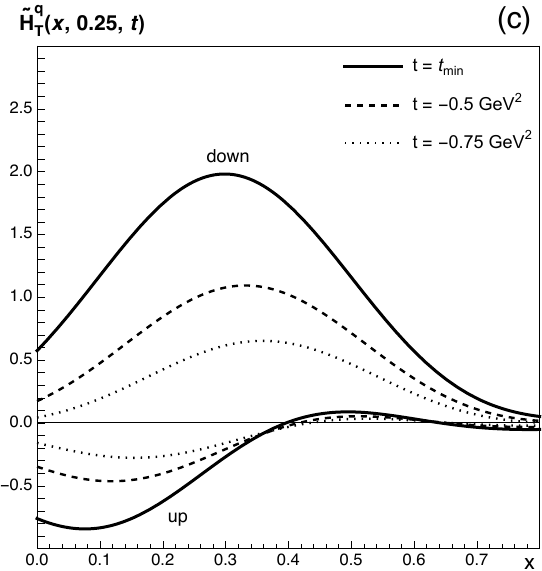} \
\includegraphics[width=4.2cm]{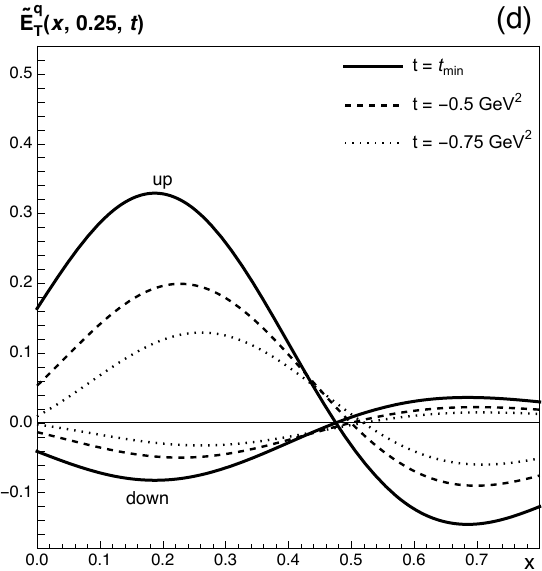} 
\par\end{centering}
\caption{\label{Fig2:ChiralOdd055-fixed-xi-various-t}
The chiral-odd GPDs 
(a) $H_T^q(x,\xi,t)$,
(b) $E_T^q(x,\xi,t)$,
(c) $\tilde{H}^q_T(x,\xi,t)$, 
(d) $\tilde{E}^q_T(x,\xi,t)$ in bag model for a fixed value of $\xi=0.25$
and selected values of $t$ starting from 
$t_{\rm min} = -0.235 \, \rm GeV^2$ over
$t = -0.5 \, \rm GeV^2$ to
$t = -0.75\, \rm GeV^2$.}
\end{figure}

In Fig.~\ref{Fig2:ChiralOdd055-fixed-xi-various-t} we illustrate
the $t$-dependence of the chiral odd GPDs. For that we fix 
$\xi=0.25$ and plot the GPDs for selected roughly 
equidistant values of $t$ starting from 
$t_{\rm min} = -0.235 \, \rm GeV^2$ over
$t = -0.5 \, \rm GeV^2$ to
$t = -0.75\, \rm GeV^2$.
Note that $t_{\rm min}$ is the minimal value of $(-t)$
for fixed $\xi$, i.e.\ the components $\Delta^i_\perp$ 
are zero and the entire momentum transfer to
the nucleon is along the $z$-axis chosen for the lightcone 
direction. In all cases in 
Fig.~\ref{Fig2:ChiralOdd055-fixed-xi-various-t} we observe
that the magnitude of the GPDs strongly decreases as
$|t|$ increases. Noteworthy is that the
position of the nodes of $\tilde{H}_T^u(x,\xi,t)$ and
$\tilde{E}_T^q(x,\xi,t)$ for $q=u,\,d$ is
relatively stable as $t$ varies. 


\begin{wrapfigure}[15]{tr}{7cm}
\vspace{-5mm}
\begin{centering}
\includegraphics[width=4.5cm]{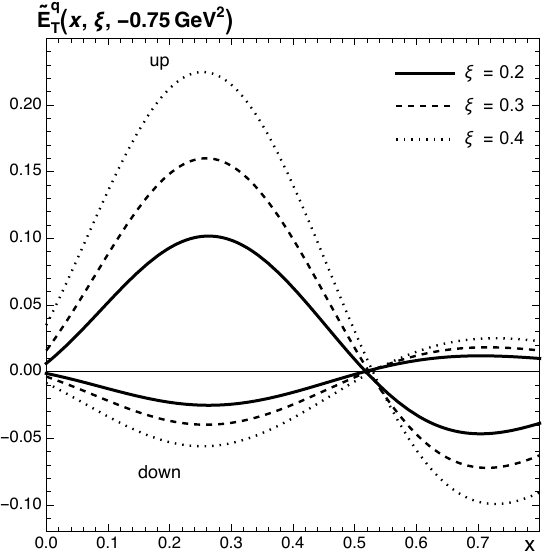}
\par\end{centering}
\caption{\label{Fig3:ETtilde055-fixed-t-varios-xi}
The GPD $\tilde{E}_T^q(x,\xi,t)$ in the bag model for $t=-0.75\,\rm GeV^2$ and selected 
 $\xi=0.2,\,0.3,\,0.4$.}
\end{wrapfigure}

Next we discuss the $\xi$-dependence of chiral-odd GPDs 
for fixed~$t$.  In the bag model, the GPDs
$H_T^q(x,\xi,t)$, $E_T^q(x,\xi,t)$, $\tilde{H}^q_T(x,\xi,t)$
show so little $\xi$-dependence that curves referring to a common
value of $t$ and various $\xi$ would be hardly distinguishable
in the figures, and we refrain from showing them. The same 
observation was made in Ref.~\cite{Ji:1997gm} for 
chiral-even GPDs in the bag model. 
We are not aware of an explanation for this bag-model
specific 
feature. In other models, one may 
observe stronger $\xi$-dependencies, see 
Sec.~\ref{Sec-6:compare-to-models}.

The situation is different for $\tilde{E}^q_T(x,\xi,t)$ which differs
from all other GPDs in one important respect.
The three chiral-odd GPDs $H_T^q(x,\xi,t)$, $E_T^q(x,\xi,t)$, 
$\tilde{H}^q_T(x,\xi,t)$ (and all chiral-even GPDs) are even functions 
of $\xi$, see Eq.~(\ref{skewness1}). In contrast to this, 
$\tilde{E}^q_T(x,\xi,t)$ is an odd function of $\xi$ according 
to Eq.~(\ref{skewness2}), and as such it is also the only GPD in the
bag model to exhibit a worthwhile mentioning $\xi$-dependence.
In fact, for $\xi=0$ it vanishes and for $\xi>0$, to a good approximation, 
$\tilde{E}^q_T(x,\xi,t)$ is linearly increasing with $\xi$ until reaching 
the largest kinematically possible value of $\xi$ for a given $t$. 
In Fig.~\ref{Fig3:ETtilde055-fixed-t-varios-xi} we illustrate 
this for $t=-0.75\,{\rm GeV}^2$ and selected values of 
$\xi$. Noteworthy is that the position of the node of 
$\tilde{E}^q_T(x,\xi,t)$ for a fixed value of $t$ in
Fig.~\ref{Fig3:ETtilde055-fixed-t-varios-xi} is unaffected as $\xi$ 
is varied. 

\begin{figure}[h!]
\begin{centering}
\includegraphics[width=4.2cm]{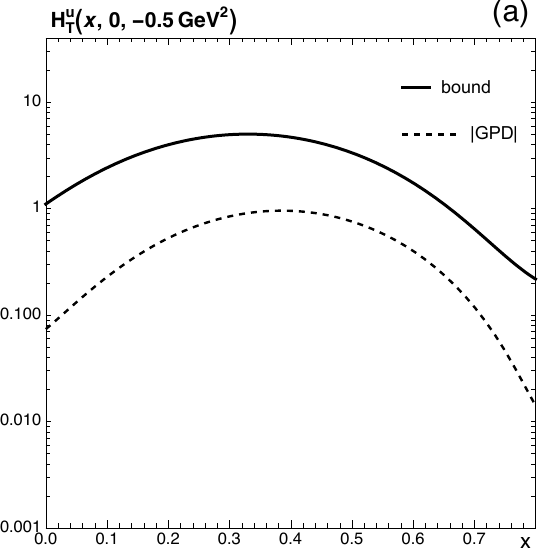} \
\includegraphics[width=4.2cm]{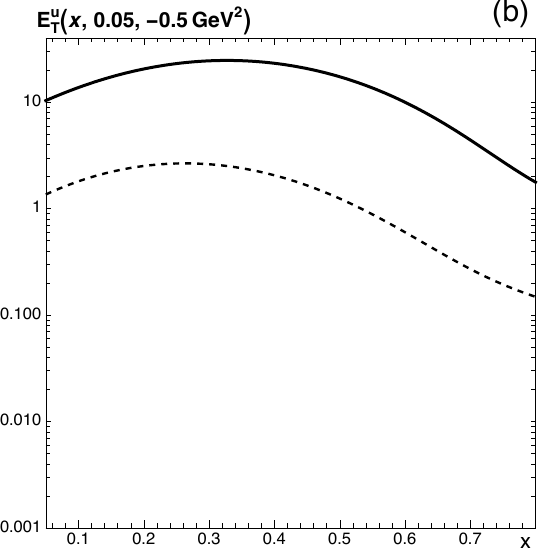} \
\includegraphics[width=4.2cm]{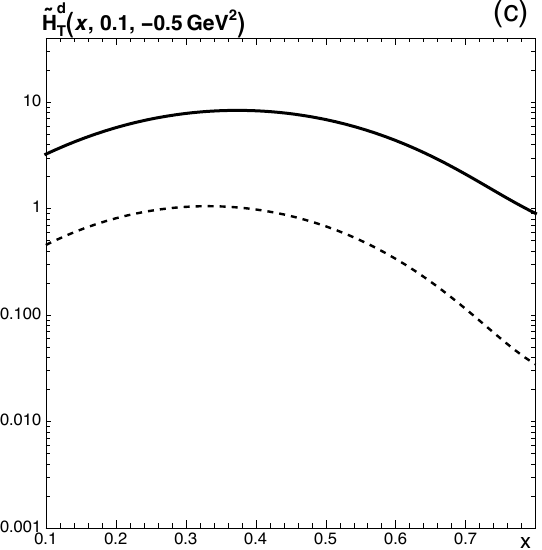} \
\includegraphics[width=4.2cm]{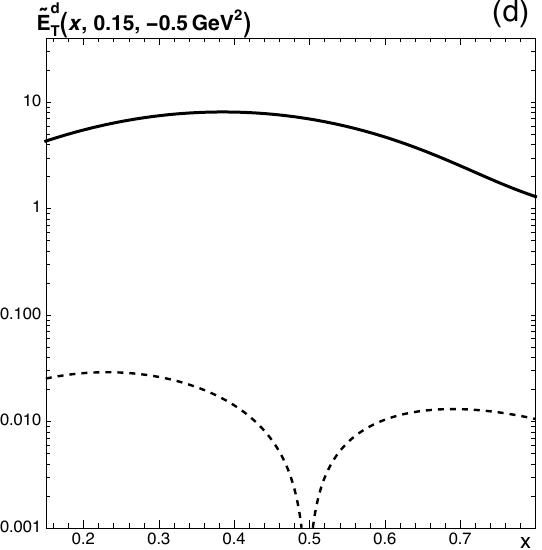} 
\par\end{centering}
\caption{\label{Fig7:ChiralOddBounds}
Test of the inequalities (\ref{ineq:HT}-\ref{ineq:ildeHT})
for the chiral-odd GPDs  in the bag model at a common value of 
$t=-0.5\,{\rm GeV}^2$ and various values of $\xi$ for the GPDs
(a) $H_T^u(x,0,t)$,
(b) $E_T^u(x,0.05,t)$,
(c) $\tilde{H}^d_T(x,0.1,t)$, 
(d) $\tilde{E}^d_T(x,0.15,t)$.
Solid lines: upper bounds for the GPDs as given by the 
right-hand-sides in (\ref{ineq:HT}-\ref{ineq:ildeHT}).
Dashed lines: the absolute values of the respective GPDs.}
\end{figure}

We end this section with an investigation of the inequalities for chiral-odd GPDs
in Eqs.~(\ref{ineq:HT}-\ref{ineq:ildeHT}) which provides an important consistency
test for the approach. The inequalities are valid in the bag model. We illustrate 
this representatively for selected flavors and values of $\xi$ and $t$ in 
Fig.~\ref{Fig7:ChiralOddBounds}. 
In this figure, the solid lines show the upper bounds for the GPDs, i.e.\ 
respectively the right-hand-sides in Eqs.~(\ref{ineq:HT}-\ref{ineq:ildeHT}),
which are defined in terms of the PDFs $f_1^q(x)$, $g_1^q(x)$, $h_1^q(x)$
evaluated in the bag model.
The dashed lines in Fig.~\ref{Fig7:ChiralOddBounds} show the respective 
absolute values of the pertinent GPDs. In Fig.~\ref{Fig7:ChiralOddBounds} 
we have chosen a common value of $t=-0.5\,{\rm GeV}^2$ for all four chiral-odd
GPDs, while we have chosen $\xi=0,\,0.05,\,0.1,\,0.15$ for respectively  
$H_T^u(x,\xi,t)$ in Fig.~\ref{Fig7:ChiralOddBounds}a,
$E_T^u(x,\xi,t)$ in Fig.~\ref{Fig7:ChiralOddBounds}b,
$\tilde{H}^d_T(x,\xi,t)$ in Fig.~\ref{Fig7:ChiralOddBounds}c, and
$\tilde{E}^d_T(x,\xi,t)$  in Fig.~\ref{Fig7:ChiralOddBounds}d.
Notice that Figs.~\ref{Fig7:ChiralOddBounds}a and \ref{Fig7:ChiralOddBounds}b
show GPDs of $u$-flavor, while 
Figs.~\ref{Fig7:ChiralOddBounds}c and \ref{Fig7:ChiralOddBounds}d
show GPDs of $d$-flavor.
In all cases, the bounds are significantly larger than the GPDs such that a
logarithmic scale on the y-axis is convenient for display.
The GPDs $H_T^u(x,\xi,t)$, $E_T^u(x,\xi,t)$, $\tilde{H}^d_T(x,\xi,t)$ stay below 
their respective bounds by almost 1 order of magnitude in 
Figs.~\ref{Fig7:ChiralOddBounds}a-c, while $\tilde{E}_T^q(x,\xi,t)$ stays below 
its bound by about 2 orders of magnitude in Fig.~\ref{Fig7:ChiralOddBounds}d.

\newpage

\section{Comparison to quark model results in literature}
\label{Sec-6:compare-to-models}

The first model study of chiral odd GPDs in the bag model 
was carried out in Ref.~\cite{Scopetta:2005fg} following the
methods of Ref.~\cite{Ji:1997gm} as we did, but a different 
result was obtained, namely only $H_T^q(x,\xi,t)$ was argued 
to be non-zero while other chiral-odd GPDs 
were incorrectly assumed to vanish. The incorrect conclusions of 
Ref.~\cite{Scopetta:2005fg} can be traced back to two steps. 
In a first (incomplete) step, only a certain linear combination 
of the structures ${\cal M}_{\lambdaS^\prime\lambdaS}[\Gamma]$ 
in Eq.~(\ref{Eq:abbreviation}) was explored, namely 
${\cal M}_{\lambdaS^\prime\lambdaS}[\sigma^{+1}+i\sigma^{+2}]$.
This combination has a certain interpretation in terms of nucleon-quark helicity amplitudes \cite{Diehl:2001pm}. 
In each of the four amplitudes describing the chiral-odd case
the quark spin flips with various possibilities for the nucleon spin to flip or 
not to flip. This first step in Ref.~\cite{Scopetta:2005fg} was incomplete because 
the chiral-odd GPDs are generally defined in terms of the structures 
${\cal M}_{\lambdaS^\prime\lambdaS}[\sigma^{+j}]$ in Eq.~(\ref{oddGPDs}). 
Indeed, in our derivation it was beneficial to make
use of the freedom to choose freely both values for the transverse 
index $j=1,\,2$, cf.~Eqs.~(\ref{T1}-\ref{T4}). In the second (incorrect) 
step it was assumed that the nucleon spin cannot flip in the bag model
because it was assumed to be associated with the bag model quark wave
functions. This step is incorrect because the necessary units of orbital
angular momentum which compensate the nucleon spin-flip in the 
nucleon-quark helicity amplitudes are provided by the explicit
factors of $\Delta^\mu$ in the decomposition in Eq.~(\ref{oddGPDs}). 
As a consequence of these two steps, only one of the possible spin structures 
of ${\cal M}_{\lambdaS^\prime\lambdaS}[\Gamma]$ was explored in
combination with only one choice $\Gamma=\sigma^{+1}+i\sigma^{+2}$
leading to only one equation which, under the assumption that the
other chiral-odd GPDs vanish, was thought to describe $H_T^q(x,\xi,t)$
while it actually describes a certain linear combination of chiral-odd
GPDs.

\begin{figure}[b!]
\begin{centering}
\includegraphics[width=4.2cm]{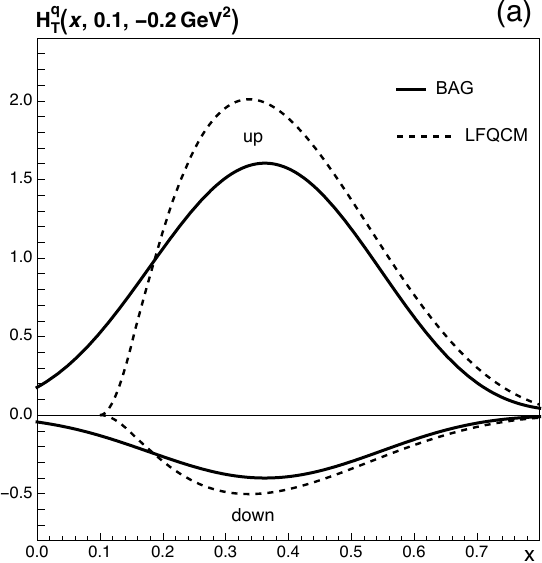} \
\includegraphics[width=4.2cm]{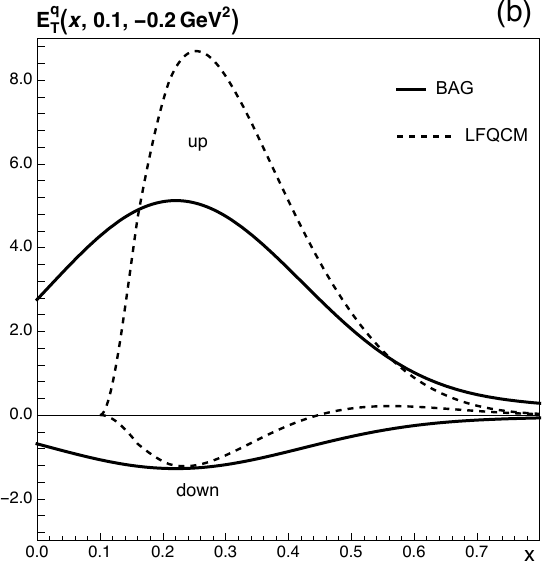} \
\includegraphics[width=4.2cm]{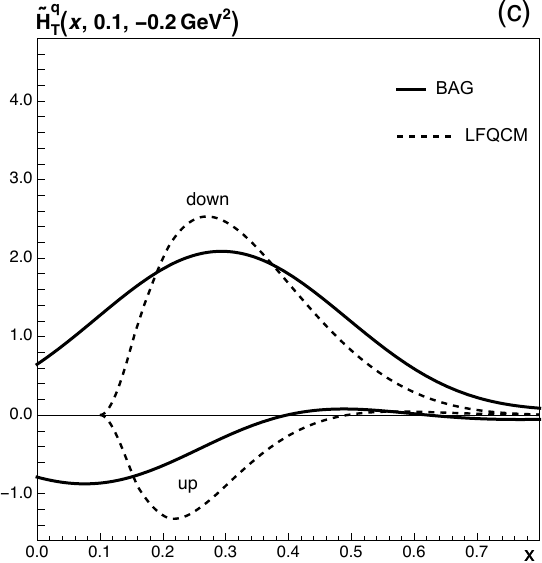} \
\includegraphics[width=4.2cm]{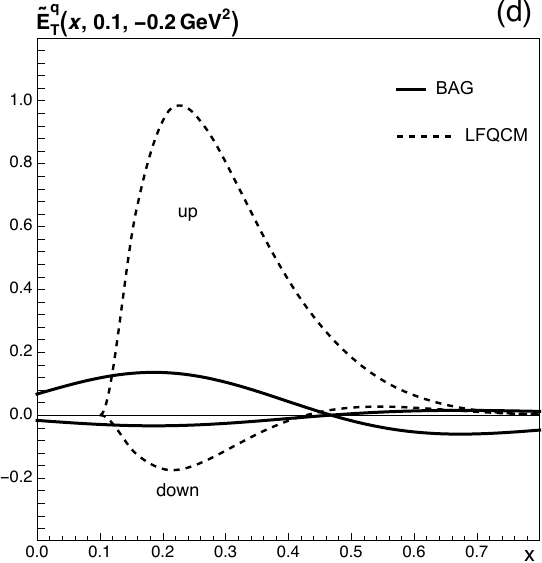}
\par\end{centering}
\caption{\label{Fig4:compare-BAG-LFCQM}
The chiral-odd GPDs 
(a) $H_T^q(x,\xi,t)$,
(b) $E_T^q(x,\xi,t)$,
(c) $\tilde{H}^q_T(x,\xi,t)$, 
(d) $\tilde{E}^q_T(x,\xi,t)$ 
at $\xi=0.1$ and $t=-0.2\,\rm GeV^2$ based on the 
bag model (solid lines, this work)
and the LFCQM (dashed lines, Ref.~\cite{Lorce:2011dv}).}
\end{figure}

\begin{figure}[b!]
\begin{centering}
\includegraphics[width=4cm]{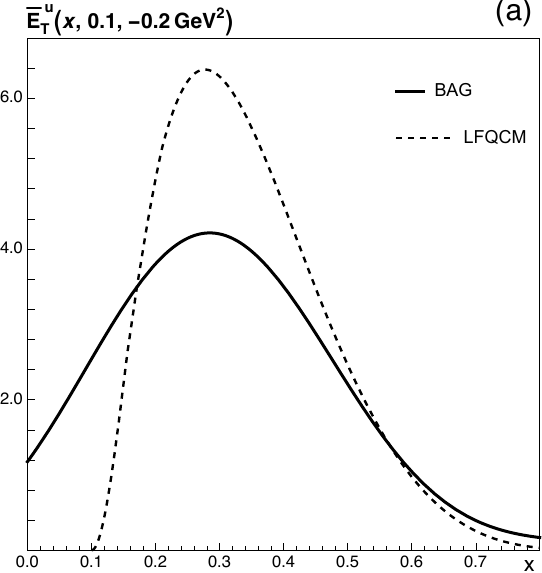} \
\includegraphics[width=4cm]{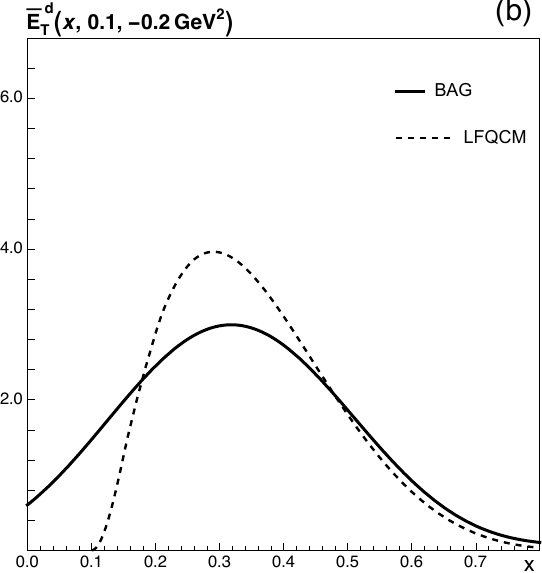}
\par\end{centering}
\caption{\label{Fig5:compare-BAG-LFCQM-barE}
The chiral-odd GPDs 
(a) $\overline{E}_T^u(x,\xi,t)$,
(b) $\overline{E}_T^d(x,\xi,t)$
at $\xi=0.1$ and $t=-0.2\,\rm GeV^2$ based on the 
bag model (solid lines, this work)
and the LFCQM (dashed lines, Ref.~\cite{Lorce:2011dv}).}
\end{figure}

To the best of our knowledge, the first model study of all four chiral 
odd GPDs was carried out in the Light-Front Constituent Quark Model (LFCQM) 
\cite{Pasquini:2005dk}, see also \cite{Pasquini:2007xz,Lorce:2011dv},
where nucleon light-front wave functions are modelled in a Fock space expansion 
truncated after the lowest $qqq$-Fock component. To this order 
in the Fock expansion, it is possible to model GPDs only in the region $x>\xi$ 
(modelling for $x<\xi$ requires at least a  $qqqq\bar q$-Fock component).
The underlying wave functions can be taken from different models
\cite{Schlumpf:1992pp,Schmidt:1997vm,Faccioli:1998aq}.
In Fig.~\ref{Fig4:compare-BAG-LFCQM} we compare the bag model predictions for
$H_T^q(x,\xi,t)$,
$E_T^q(x,\xi,t)$,
$\tilde{H}^q_T(x,\xi,t)$, 
$\tilde{E}^q_T(x,\xi,t)$  at $\xi=0.1$ and $t=-0.2\,\rm GeV^2$
with LFCQM results \cite{Lorce:2011dv} from nucleon light-front wave functions 
constructed on the basis of the relativistic quark model \cite{Schlumpf:1992pp}.
The predictions from both models refer to a comparably low scale 
\cite{Stratmann:1993aw,Boffi:2009sh,Pasquini:2011tk,Pasquini:2014ppa}.
Considering that in the LFCQM  the GPDs vanish for $x<\xi$,
in our comparison we focus on $x\gtrsim 0.2$. In this region, we 
observe in Fig.~\ref{Fig4:compare-BAG-LFCQM}a very good agreement 
for $H_T^q(x,\xi,t)$ for $u$ and $d$ flavors within $20\,\%$ 
or better. In Fig.~\ref{Fig4:compare-BAG-LFCQM}b we find for 
$E_T^u(x,\xi,t)$ a somewhat wider spread, but the results still agree 
within about $40\,\%$, while for the $d$-flavor the two models are in 
qualitative agreement in the sense that $E_T^d(x,\xi,t)$ is much smaller 
than the corresponding $u$-quark GPD, although $E_T^d(x,\xi,t)$ 
exhibits a node in LFCQM not seen in bag model. 
The two models also agree on the flavor dependence in the case of 
$\tilde{H}_T^q(x,\xi,t)$ with good numerical agreement on the
dominant $d$ flavor in the valence-$x$ region and qualitative agreement on 
the smallness of the $u$-flavor, see Fig.~\ref{Fig4:compare-BAG-LFCQM}c. 
Merely in the case of  $\tilde{E}_T^q(x,\xi,t)$ in 
Fig.~\ref{Fig4:compare-BAG-LFCQM}d, we observe no good agreement, with
the bag model result being substantially smaller than the LFCQM result.
This discrepancy of the
model predictions in the case of $\tilde{E}_T^q(x,\xi,t)$ is not
surprising. This GPD has particular properties which distinguish it
from all other GPDs. In fact, in an approach where GPDs can only be 
modelled for $x>\xi$, it is not possible to satisfy the property
$\tilde{E}_T^q(x,-\xi,t)=-\tilde{E}_T^q(x,\xi,t)$. As can be seen in 
Fig.~\ref{Fig4:compare-BAG-LFCQM}d, the LFCQM based on the minimal
$qqq$-Fock component also cannot comply with the constraint 
$\int dx\,\tilde{E}_T^q(x,\xi,t)=0$. 
Clearly, higher Fock components must play an important role, especially 
for the GPD $\tilde{E}_T^q(x,\xi,t)$.
For phenomenological applications, the combination 
$\overline{E}_T^q(x,\xi,t) = E_T^q(x,\xi,t)+2\,\tilde{H}_T^q(x,\xi,t)$ 
is of importance rather than the individual GPDs.
As can be seen in Fig.~\ref{Fig5:compare-BAG-LFCQM-barE}, in both models these GPDs have 
positive signs and comparable magnitudes within about 20-40$\,\%$. The same sign-result 
for $u$- and $d$-flavor of $\overline{E}_T^q(x,\xi,t)$ is remarkable because the individual 
chiral-odd GPDs entering $\overline{E}_T^q(x,\xi,t)$ 
generally follow the pattern of opposite signs for $u$- and $d$-flavors.

In the chiral quark soliton model \cite{Wakamatsu:2008ki}, the chiral-odd GPDs 
with a non-trivial forward limit were studied, and three linearly independent equations 
were derived for them.  Due to 
``quite involved'' numerics, only one linear combination,
namely $G_T^q(x,0,0)=H_T^q(x,0,0)+E_T^q(x,0,0)+ 2\tilde{H}_T^q(x,0,0)$,
was computed.
The chiral quark soliton model is based on the $1/N_c$ expansion and it is 
convenient to compute results for the flavor combinations $(u\pm d)$.
The bag and chiral quark soliton model agree that the isoscalar flavor combination 
$G_T^{u+d}(x,0,0)$ is sizeable, and the isovector flavor combination 
$G_T^{u-d}(x,0,0)$ is somewhat smaller.
For the latter, we find good quantitative agreement in the two models within about 
10-20$\,\%$ accuracy. For $G_T^{u+d}(x,0,0)$, the two models agree on the sign and 
magnitude for $x > 0.1$ with the chiral quark soliton result somewhat shifted towards 
smaller~$x$. The small-$x$ behavior of $G_T^{u-d}(x,0,0)$ is qualitatively different in the two models,
but one needs to keep in mind that quark models are generally not applicable for $x< 0.1$.
The chiral quark soliton model can also be used to construct input nucleon light-cone
wave functions for the LFCQM approach with results similar to the LFCQM results shown 
in Fig.~\ref{Fig4:compare-BAG-LFCQM} \cite{Lorce:2011dv}.

In the scalar quark-diquark model based on light-front wave functions obtained in a 
soft-wall AdS/QCD correspondence approach \cite{Chakrabarti:2015ama,Maji:2017ill},
the GPDs were modelled based on the overlap representation in terms
of the minimal quark-diquark Fock component. Similarly to the above-discussed LFCQM, 
this approach can describe GPDs only for $x>\xi$ and cannot comply with the sum rule 
(\ref{odd-first-moment}) for $\tilde{E}_T^q(x,\xi,t)$. Except for the latter, our results
for chiral-odd GPDs are in agreement with \cite{Chakrabarti:2015ama,Maji:2017ill} 
within the accuracy one can expect from quark models. 

In Ref.~\cite{Kaur:2023lun} the chiral-odd GPDs 
$H_T^q(x,0,t)$, $E_T^q(x,0,t)$, $\tilde{H}_T^q(x,0,t)$ where studied in the limit 
$\xi\to 0$ (we recall that  $\tilde{E}_T^q(x,\xi,t)\to 0$ in this limit) in the
basis light-front quantization approach with an effective light-front Hamiltonian 
which incorporates confinement and one-gluon exchange between valence quarks.
Except for the region $x<0.1$ (where quark model results cannot be expected to be
reliable), we observe a good agreement with the results from  Ref.~\cite{Kaur:2023lun}.
This model was recently extended to $\xi\neq0$ observing a stronger
$\xi$-dependence \cite{Liu:2024umn} as compared to the bag model.

In the reggeized diquark model in \cite{Goldstein:2013gra} physically motivated parameterizations
of chiral-odd GPDs were proposed. The results shown in  \cite{Goldstein:2013gra} refer to a scale 
of $2\,\rm GeV^2$ and cannot be directly compared to our results which refer to a low model scale. 
The reggeized diquark model was constructed with polynomiality imposed as a constraint
\cite{Goldstein:2013gra}. For studies of chiral-odd GPDs in models based on generalized QED,
we refer to Refs.~\cite{Dahiya:2007mt,Chakrabarti:2008mw}.

\section{\boldmath Comparison to predictions from the large-$N_c$ limit in QCD}
\label{Sec-7:compare-to-large-Nc}

The bag model results are in agreement with the flavor dependence of the GPDs 
$H_T^q(x,\xi,t)$, $\tilde{E}_T^q(x,\xi,t)$, $\overline{E}_T^q(x,\xi,t)$ predicted in the 
large $N_c$ limit, where the $(u\pm d)$ flavor combinations 
exhibit different $N_c$-behaviors.
The $(u-d)$ flavor combinations of $H_T^q(x,\xi,t)$ and 
$\tilde{E}_T^q(x,\xi,t)$ are one order in $N_c$ enhanced over the respective 
$(u+d)$ flavor combinations, while for $\overline{E}_T^q(x,\xi,t)$ the situation is reversed. 
In the strict limit $N_c\to\infty$ one has \cite{Schweitzer:2016jmd,Schweitzer:2015zxa}
\be\label{Eq:large-Nc}
    H_T^u(x,\xi,t)          \approx -          H_T^d(x,\xi,t) \sim N_c^2 , \quad
    \tilde{E}_T^u(x,\xi,t)  \approx - \tilde{E}_T^d(x,\xi,t)  \sim N_c^3 , \quad
    \overline{E}_T^u(x,\xi,t)    \approx     \overline{E}_T^d(x,\xi,t)  \sim N_c^3  .
\ee
Our results are compatible with the large-$N_c$ predictions
(\ref{Eq:large-Nc}) in the sense that the $u$- and $d$-flavors of $H_T^q(x,\xi,t)$ 
and $\tilde{E}_T^q(x,\xi,t)$ exhibit opposite signs and those of $\overline{E}_T^q(x,\xi,t)$
exhibit the same signs, see  Figs.~\ref{Fig4:compare-BAG-LFCQM} and \ref{Fig5:compare-BAG-LFCQM-barE}.

Interestingly, the $N_c$-dependence of the individual GPDs entering $\overline{E}_T^d(x,\xi,t)$, 
namely $E_T^q(x,\xi,t)$ and $\tilde{H}_T^q(x,\xi,t)$, differs from that obtained in 
Refs.~\cite{Schweitzer:2016jmd,Schweitzer:2015zxa}. 
This topic deserves further investigations.

\begin{figure}[b!]
\vspace{3mm}
\begin{centering}
\includegraphics[height=5.5cm]{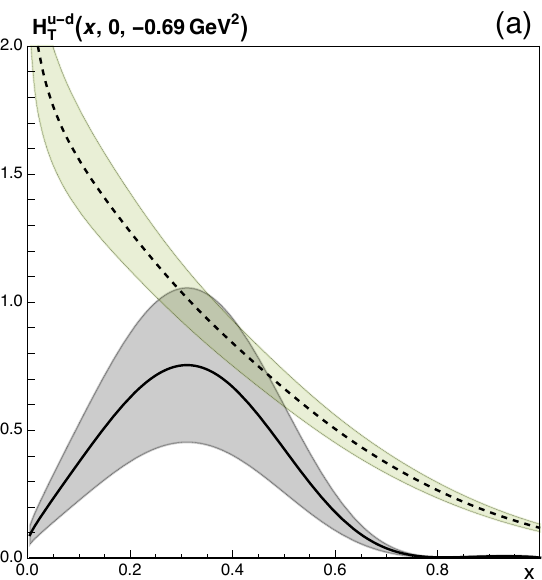} \
\includegraphics[height=5.5cm]{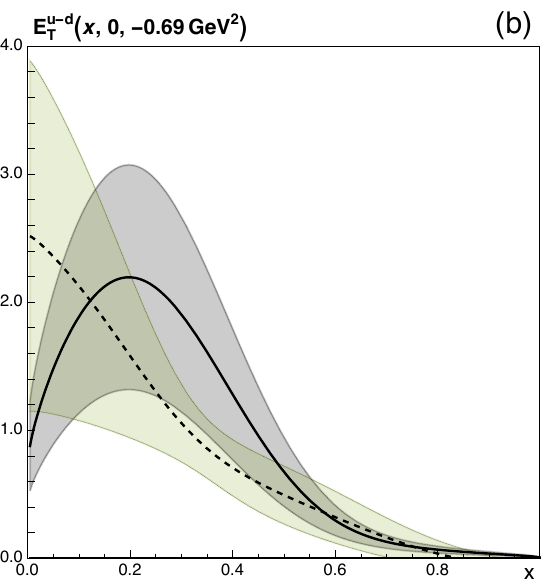} \
\includegraphics[height=5.5cm]{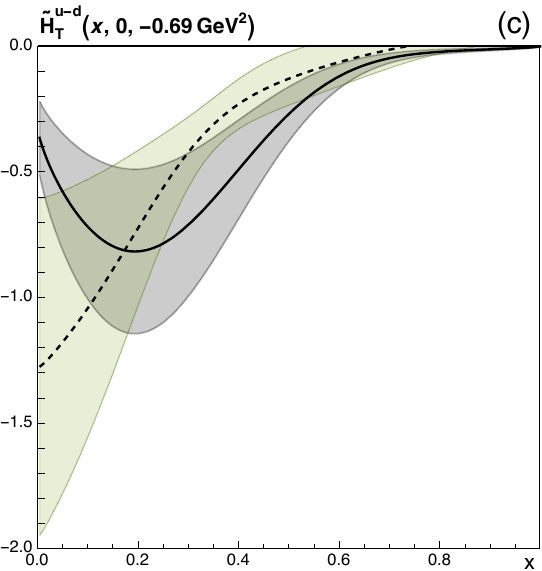} \
\par\end{centering}
\caption{\label{Fig6:compare-lattice}
Chiral-odd GPDs 
(a) $H_T^{u-d}(x,0,t)$,
(b) $E_T^{u-d}(x,0,t)$,
(c) $\tilde{H}^{u-d}_T(x,0,t)$ from the bag model (this work, solid lines)
in comparison to the corresponding quasi GPDs from lattice QCD 
(Ref.~\cite{Alexandrou:2021bbo}, dashed lines) for 
$P_z=1.67\,\rm GeV$ in $\overline{\rm MS}$ scheme. All results 
refer to $t=-0.69\,{\rm GeV}^2$ and the scale $\mu=2\,{\rm GeV}$.
The shaded areas show estimated theoretical uncertainties, see text.}
\end{figure} 

\section{Comparison to lattice QCD calculations of quasi GPDs}
\label{Sec-8:compare-to-lattice}

Chiral-odd GPDs have been studied in lattice QCD in 
Ref.~\cite{Alexandrou:2021bbo} by exploring the quasi distribution method~\cite{Ji:2013dva}. 
In this method, one considers nonlocal operators similar to those in Eq.~(\ref{Eq:abbreviation}) 
except that the field operators have a spacelike separation, e.g.\ \ $z^\mu=(0,0,0,z^3)$,
which can be implemented on a Euclidean lattice, in contrast to lightlike separations 
in the actual GPD definition in Eq.~(\ref{Eq:abbreviation}).
Sandwiching such an operator for an adequately chosen Dirac matrix $\Gamma$ between 
nucleon states with momenta $\vec{p}=(0\,,0\,,P_z)-\vec{\Delta}/2$ and 
$\vec{p}^{\,\prime}=(0\,,0\,,P_z)+\vec{\Delta}/2$, one can compute on a Euclidean lattice 
so-called quasi GPDs, i.e.\ $H^q_T(x,\xi,t,P_z)$ and so on, which
are frame-dependent and reduce to regular GPDs in the limit $P_z\to\infty$ 
(for brevity we do not indicate the renormalization scale dependence).
The exact connection of quasi and lightcone distributions (for PDFs, GPDs)
in QCD is nontrivial (since the operations of (i) renormalizing the operators and (ii) taking 
the limit $P_z\to\infty$ do not commute) and handled in the formalism of the large-momentum 
effective theory, see Refs.~\cite{Ji:2020ect,Constantinou:2020pek,Constantinou:2022yye} 
for reviews. One of the obvious practical limitations of the quasi distribution method is 
to reach large $P_z$ in a lattice QCD calculation and carry out the limit $P_z\to\infty$ 
with systematic uncertainties under control. This is still difficult 
to carry out at the current state. 

In Ref.~\cite{Alexandrou:2021bbo} the $(u-d)$ flavor combinations were computed which 
has the advantage in lattice QCD calculations that disconnected diagrams do not contribute.
In Ref.~\cite{Alexandrou:2021bbo} results were obtained for $(-t)\le 0.69\,\rm GeV^2$, 
$\xi=0$ or $1/3$, and $P_z$ up to 1.67$\,$GeV and the limit $P_z\to\infty$ was not taken.
A direct comparison of our results to the lattice data is not possible for two reasons. 
First, the lattice results refer to a renormalization scale $\mu = 2\,\rm GeV$ 
(in $\overline{\rm MS}$ scheme) while the model results refer to a low scale. 
Second, the model results correspond to GPDs while the lattice results correspond 
to quasi GPDs for $P_z \le 1.67\,$GeV. To present a tentative comparison 
to the lattice QCD results, we shall make an attempt to remedy the first point.
The second point cannot be addressed at this time and must be postponed until 
lattice data at sufficiently large $P_z$ become available to make an extrapolation 
$P_z\to\infty$ feasible.

We choose to compare the model and lattice results at $\xi=0$ where instead
of the more complicated GPD evolution \cite{Belitsky:2000yn}, we can simply 
use the DGLAP evolution of $h_1^q(x)$ \cite{Artru:1989zv} for the chiral-odd
GPDs except for $\tilde{E}_T^q(x,0,t)$ which is zero. (For completeness, we note 
that the one-loop evolution kernels of chiral-odd GPDs have been calculated 
independently and numerically implemented \cite{Bertone:2022frx} in the PARTONS 
framework \cite{Berthou:2015oaw}.) 
The exact value of the initial scale $\mu_0$ in a quark model is not 
well known \cite{Stratmann:1993aw,Boffi:2009sh,Pasquini:2011tk,Pasquini:2014ppa}.
In general, it can be expected that $\mu_0$ does not exceed the nucleon mass.
In models based on valence quark degrees of freedom, it was estimated 
$\mu_0\approx 0.5\,\rm GeV$ \cite{Pasquini:2014ppa}.
We therefore carry out the leading-order 
evolution varying the initial scale conservatively in a wide interval 
$0.5 \le \mu_0 \le 0.9\,{\rm GeV}$ based on the evolution code used 
in Ref.~\cite{Schweitzer:2001sr}.

In Fig.~\ref{Fig6:compare-lattice}, the solid lines show the bag model results for
$H_T^{u-d}(x,0,t)$, $E_T^{u-d}(x,0,t)$, $\tilde{H}_T^{u-d}(x,0,t)$ leading-order evolved 
from $\mu_0 = 0.7\,{\rm GeV}$ to the scale of the lattice results $\mu=2\,$GeV. 
The shaded regions around the middle values indicate
``model error bands'' of $\pm\,40\,\%$ uncertainty. This is a conservative estimate
of the model dependence in the following sense. This model error band 
(i) contains the variation of the bag model parameter $\eta=0.35$ or 0.55 in Fig.~\ref{Fig1:ChiralOddForward}, 
(ii) approximately represents the uncertainty as inferred from the comparison of
bag model and LFCQM results in Figs.~\ref{Fig4:compare-BAG-LFCQM} and 
\ref{Fig5:compare-BAG-LFCQM-barE}, and (iii) covers the variation of the initial 
scale in the interval $0.5 \lesssim \mu_0 \lesssim 0.9\,{\rm GeV}$.
(Also, the lattice results refer to next-to-leading order in $\overline{\rm MS}$ scheme.
We assume the systematic uncertainty associated with this fact to be included in our conservative 
uncertainty estimate.)

The bag model results evolved in this way along with their estimated model uncertainty
shown in Fig.~\ref{Fig6:compare-lattice} refer hence to the same scale as the lattice QCD
results from Ref.~\cite{Alexandrou:2021bbo} included in 
Fig.~\ref{Fig6:compare-lattice} along with their uncertainties. 
The uncertainty band of the lattice results includes the statistical and other theoretical 
uncertainties, but no estimate of the uncertainty associated with the extrapolation 
$P_z\to\infty$. Indeed, the lattice results refer to quasi GPDs, i.e.\ to 
$H_T^{u-d}(x,0,t,P_z)$, $E_T^{u-d}(x,0,t,P_z)$, $\tilde{H}_T^{u-d}(x,0,t,P_z)$
with $P_z = 1.67\,{\rm GeV}$, the largest value of $P_z$ used in the pioneering 
lattice calculation \cite{Alexandrou:2021bbo}. In the future, when it will be 
possible to run lattice calculations with still larger $P_z$, it will become
feasible to extrapolate to $P_z\to\infty$. Currently, this is not possible, and we
need to keep this grain of salt in mind. It is also important to keep in mind a 
generic limitation of the quasi distribution approach which is not applicable 
at small or large $x$ due to the large-momentum effective theory framework 
requiring power corrections of the type $M_{\rm non}^2/(x^2(1-x)P_z^2)$ 
to be small where $M_{\rm non}$ 
is a nonperturbative scale \cite{Ji:2020ect,Constantinou:2020pek,Constantinou:2022yye}.
These power correction are not negligible when $x$ is small or close to 1
(for $\xi\neq0$ there are additional limitations when $|x-\xi|$ becomes small).

Keeping these reservations in mind, the lattice result \cite{Alexandrou:2021bbo} 
for $H_T^{u-d}(x,0,t,P_z)$ (referring to $P_z = 1.67\,{\rm GeV}$) and the model 
result (corresponding to $P_z\to\infty$) agree qualitatively in 
Fig.~\ref{Fig6:compare-lattice}a, especially in the valence-$x$ region 
$0.2 \lesssim x\lesssim 0.5$ where quark models might be expected to work most
reliably. Noteworthy is that the lattice result for $H_T^{u-d}(x,0,t,P_z)$ in
Fig.~\ref{Fig6:compare-lattice}a has the smallest theoretical uncertainty of 
all quasi GPDs from \cite{Alexandrou:2021bbo}. 
It of course remains to be seen how the lattice result will change when larger 
values of $P_z$ become available and the limit $P_z\to\infty$ can be taken.

The lattice results for $E_T^{u-d}(x,0,t,P_z)$ and $\tilde{H}_T^{u-d}(x,0,t,P_z)$
from \cite{Alexandrou:2021bbo} shown in Figs.~\ref{Fig6:compare-lattice}b and 
\ref{Fig6:compare-lattice}c have larger uncertainties than $H_T^{u-d}(x,0,t,P_z)$ 
in Fig.~\ref{Fig6:compare-lattice}a because these functions appear with prefactors 
of $\Delta^\mu$ in Eq.~(\ref{oddGPDs}) and in lattice calculations, every power of
$\Delta^\mu$ is associated with more numerical noise and more involved numerics. 
Keeping in mind the limitations of the quasi distribution approach at small (and large) 
$x$-values, we observed good agreements of model and lattice results in both cases
within the uncertainties of the model and lattice calculations. This comparison
may need to be revisited when lattice computations with higher statistics and
for larger $P_z$ become available.

In our comparison to the lattice results in Fig.~\ref{Fig6:compare-lattice}, we 
have restricted ourselves to the special case $\xi=0$ where the GPD 
$\tilde{E}_T^q(x,\xi,t,P_z)$ vanishes. In Ref.~\cite{Alexandrou:2021bbo} the quasi 
GPDs were also computed for $\xi=\frac13$ and $\tilde{E}_T^{u-d}(x,\xi,t,P_z)$ was found
zero within statistical accuracy of the lattice simulation. This is understandable
based on the bag model results where $\tilde{E}_T^q(x,\xi,t)$ is by far the smallest
of all chiral-odd GPDs. In addition, the bag model predicts a node in 
$\tilde{E}_T^q(x,\xi,t)$ in the valence-$x$ region where the other three chiral-odd GPDs 
are largest. Our model results indicate that precise lattice calculations of the quasi 
GPD $\tilde{E}_T^q(x,\xi,t,P_z)$ will be challenging and require large statistics.

\section{Tensor Form Factors \& Polynomiality}
\label{Sec-9:FFs}

\begin{figure}[t!]
\begin{centering}
\includegraphics[width=4.5cm]{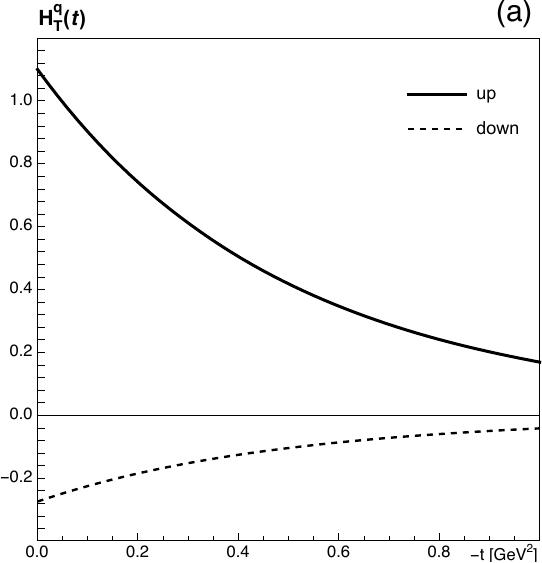} \
\includegraphics[width=4.5cm]{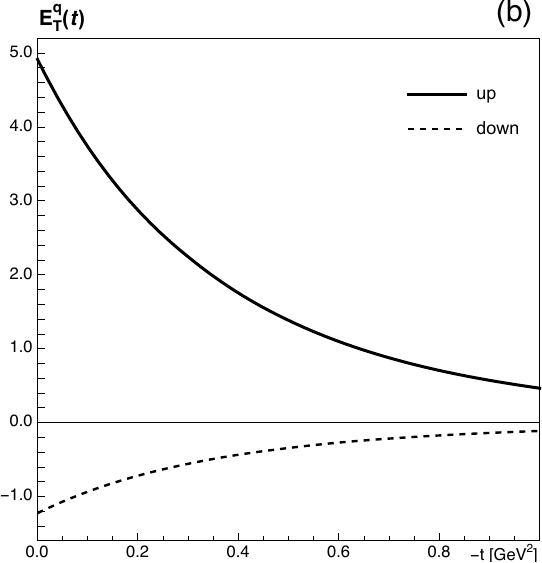} \
\includegraphics[width=4.5cm]{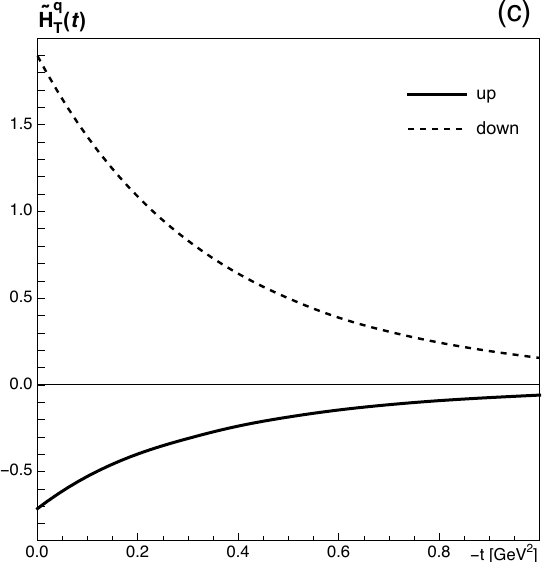}
\par\end{centering}
\caption{\label{Fig6:TFFs}
The tensor form factors 
(a) $H_T^q(t)$,
(b) $E_T^q(t)$,
(c) $\tilde{H}_T^q(t)$ in bag model for up and down quarks
at the low scale.}
\end{figure}

In this section we discuss the tensor form factors and compare to other quark models 
\cite{Pasquini:2005dk,Lorce:2011dv,Ledwig:2010zq,Gutsche:2016xff,Kaur:2023lun} and lattice QCD 
\cite{Alexandrou:2022dtc}. 
Nucleon matrix elements of the tensor current are parameterized 
in terms of three form factors
\be
\label{Eq:def-tensor-FFs}
    \la p^\prime,\vec s^{\,\prime}|  \bar{\psi}_q(0) i\sigma^{\mu\nu}\psi_q(0)|p,\vec s\rangle
    = \bar u(p^\prime,\vec s^{\,\prime})\biggl[H_T^q(t)\,i\sigma^{\mu\nu} +
    \tilde{H}_T^q(t)\, \frac{P^{\mu}\Delta^{\nu}-P^{\nu}\Delta^{\mu}}{m^2}
    + E_T^q(t)\, \frac{\gamma^{\mu}\Delta^{\nu}-\gamma^{\nu}\Delta^{\mu}}{2m}\biggl]u(p,\vec s)
    \,.
\ee 
The individual form factors in the bag model are obtained from solving 
a system of linear equations obtained from Eq.(\ref{Eq:def-tensor-FFs}) 
by exploring different $\mu, \nu$ indices and nucleon spin projections.
The procedure is analogous to the evaluation of the GPDs. The detailed
model expressions can be found in App.~\ref{App:form-factors}.
Solving the system of the three linear equations quoted in
App.~\ref{App:form-factors} in Eqs.~(\ref{Eq:FF1}-\ref{Eq:FF3})
yields the results for the tensor form factors shown in
Fig.~\ref{Fig6:TFFs}.

In the bag model, the $u:d$ ratio is exactly $4:(-1)$ for $H_T^q(t)$ and $E_T^d(t)$ and 
approximately for $\tilde{H}^q_T(t)$, see Fig.~\ref{Fig6:TFFs}. The bag model predictions 
for tensor form factors agree with other models within quark model accuracy. 
We illustrate this with a comparison in Table~\ref{Table:1} showing results, 
from the model calculations we are aware of
\cite{Pasquini:2005dk,Lorce:2011dv,Ledwig:2010zq,Gutsche:2016xff,Kaur:2023lun}, for two quantities of 
interest related to tensor form factors, namely the tensor charges and the property $\kappa_T^q$ 
defined as 
\be\label{Eq:gT-kappaT}
    g_T^q = H_T^q(0), \quad  
    \kappa_T^q = \overline{E}_T^q(0) = E_T^q(0) + 2\tilde{H}_T^q(0).
\ee
The latter has the following physical meaning: in an unpolarized nucleon moving
along the z-axis, $\kappa_T^q/(2m)$ tells us how much the average position of 
quarks polarized in the x-direction is shifted along the y-direction
\cite{Burkardt:2005hp}.

\begin{table}[b!]
\begin{tabular}{l|cccc}
\hline\hline
model (low scale $\mu_0$)                  & $g_T^u$ & $g_T^d$  & $\kappa_T^u$ & $\kappa_T^d$    \\ \hline 
bag model [this work]                                      & \ 1.10 & \ -0.28 & \ 3.49 & \ 2.57  \\
LFCQM with harmonic oscillator \cite{Pasquini:2005dk}      & \ 1.17 & \ -0.29 & \ 3.60 & \ 2.36  \\
LFCQM with hypercentral model  \cite{Pasquini:2005dk}      & \ 0.97 & \ -0.24 & \ 1.98 & \ 1.17  \\
LFCQM with Schlumpf wave func.\  \cite{Lorce:2011dv}       & \ 1.16 & \ -0.29 & \ 3.98 & \ 2.60  \\
chiral quark soliton model in SU(2) \cite{Ledwig:2010zq}   & \ 1.09 & \ -0.34 & \ 3.72 & \ 1.83 \\
chiral quark soliton model in SU(3) \cite{Ledwig:2010zq}   & \ 1.08 & \ -0.32 & \ 3.56 & \ 1.83 \\
relativistic confined quark model of \cite{Gutsche:2016xff} & \ 1.01 & \ -0.25 & \ 4.07 & \ 1.96 \\
basis light-front quantization model \cite{Kaur:2023lun} \ & \ 1.25 & \ -0.27 & \ 3.21 & \ 2.43 \\
\hline\hline
\end{tabular}
\caption{\label{Table:1} Results for the properties $g_T^q$ and $\kappa_T^q$ 
    related to tensor form factors, see Eq.~(\ref{Eq:gT-kappaT}), from different quark models: 
    bag model (this work),
    light-front constituent quark model (LFCQM) \cite{Pasquini:2005dk,Lorce:2011dv},
    chiral quark soliton model \cite{Ledwig:2010zq}, 
    relativistic confined quark model \cite{Gutsche:2016xff},
    and basis light-front quantization model \cite{Kaur:2023lun}. 
    The results refer to the low initial scales of the models.}
\end{table}

The bag model results agree well with those from the 
SU(2) and SU(3) chiral quark soliton model studies \cite{Ledwig:2010zq}, 
relativistic confined quark model of \cite{Gutsche:2016xff},
basis light-front quantization model \cite{Kaur:2023lun}, and light-front constituent 
quark model \cite{Pasquini:2005dk,Lorce:2011dv} based on respectively harmonic oscillator 
wave-functions \cite{Schmidt:1997vm} and the wave functions by Schlumpf \cite{Schlumpf:1992pp}. 
Merely the results for $\kappa_T$ from \cite{Pasquini:2005dk} 
based on the hypercentral model \cite{Faccioli:1998aq} agree to a lesser extent. 
Averaging the model results in Table~\ref{Table:1} with equal weight yields
\ba
       g_T^u &=& \phantom{-}1.10 \pm 0.09 \quad \;(8\%) \nonumber\\
       g_T^d &=&          - 0.29 \pm 0.03 \quad  (12\%) \nonumber\\
  \kappa_T^u &=& \phantom{-}3.45 \pm 0.65 \quad  (19\%) \nonumber\\
  \kappa_T^d &=& \phantom{-}2.09 \pm 0.49 \quad  (23\%) \label{Eq:models-gT-kT} 
\ea
with $\pm\,$uncertainties representing standard deviations, and the percentages
in brackets indicating relative accuracy. The models vary less for $u$- and more 
for $d$-flavor, show a lesser spread for tensor charges and larger spreads for 
$\kappa_T^q$. 

We refrain from a detailed comparison of the model predictions for the $t$-dependencies of the form
factors. Let us merely remark that in some models \cite{Ledwig:2010zq,Gutsche:2016xff,Kaur:2023lun}
the $t$-dependence of the form factors can be different for different flavors, while in the bag
model it is exactly $H_T^u(t):H_T^d(t)=4:(-1)$ and analogously for $E_T^q(t)$, and to
a good approximation also for $\tilde{H}^q_T(t)$.
Overall, the model results in Table~\ref{Table:1} as well as the results 
for the form factors for $(-t)< 1 \,\rm GeV^2$ are in line with a quark model 
accuracy of 30-40$\%$ as inferred from comparisons of GPD model calculations in
Sec.~\ref{Sec-6:compare-to-models} or transverse momentum dependent PDFs 
in Ref.~\cite{Boussarie:2023izj}. For completeness, we remark that 
in the SU(3) chiral quark soliton model the strangeness contributions $g_T^s \approx - 0.01$ 
and $\kappa_T^s \approx -0.2$ \cite{Ledwig:2010zq} were obtained.
Notice that some of the works quoted in Table~\ref{Table:1} were preceded in literature. 
For instance, in the bag model $g_T^q$ was derived in \cite{Jaffe:1991kp} and 
$\kappa_T^q$ in \cite{Burkardt:2007xm}, the light-front constituent quark model results for 
$g_T^q$ from harmonic oscillator wave functions were obtained previously in \cite{Schmidt:1997vm}, 
and $g_T^q$ in chiral quark soliton model was studied in
\cite{Kim:1995bq,Kim:1996vk,Lorce:2007fa,Ledwig:2010tu}.

In Table~\ref{Table:2} we compare the quark model results and lattice QCD calculations 
\cite{Alexandrou:2022dtc} for the properties $g_T^{u-d}$ and $\kappa_T^{u-d}$. 
We recall that in lattice QCD the computation of the isovector $(u-d)$ flavor combinations 
has the advantage that the contributions from disconnected diagrams cancel out which are
numerically expensive to determine.
The tensor current $\bar{\psi}_q\,i\sigma^{\mu\nu}\,\psi_q$ is not conserved
and the tensor charges and $\kappa_T^q$ are scale dependent. The model results 
in Table~\ref{Table:2} correspond to the values in Eq.~(\ref{Eq:models-gT-kT}) 
evolved in leading order to the scale $\mu = 2\,\rm GeV$,
cf.~Sec.~\ref{Sec-8:compare-to-lattice}.
The model dependencies are roughly estimated to be $20\%$ for $g_T^{u-d}$ and 
$30\%$ for $\kappa_T^{u-d}$ which includes the uncertainties from Table~\ref{Table:1} 
and uncertainties due to evolution from the not-well-known model scales.  
The lattice results \cite{Alexandrou:2022dtc} refer to the same scale in 
$\overline{\rm MS}$ scheme (the estimated model uncertainties are 
presumably conservative enough to include scheme dependence).
The numbers in brackets of the lattice results represent the 
uncertainties of the lattice calculation~\cite{Alexandrou:2022dtc}.

The lattice QCD results are about 2-3 times more precise than the quark model results.
This is not surprising as lattice QCD is a first principle approach where the theoretical
uncertainties are under control and can be systematically improved. In contrast to this, 
one must must not forget that the inferred model uncertainties are rough estimates.
Keeping this in mind, we observe a good agreement of the quark models and lattice QCD.
It is fair to say that the quark models successfully catch the gross features of the
properties related to tensor form factors of the nucleon.
For earlier lattice studies of tensor charges, we refer to 
\cite{Alexandrou:2019ali,QCDSF:2006tkx,Gockeler:2005cj,Aoki:1996pi}.

\begin{table}[t!]
\begin{tabular}{l|ll}
\hline\hline
approach (at $\mu = 2\,\rm GeV$)            & \ \ $g_T^{u-d}$ & \ \ $\kappa_T^{u-d}$  \\ \hline 
(i) quark models [Table~\ref{Table:1}] \    & \   $1.11(22)$  & \   $1.08(29)$  \\ 
(ii) lattice QCD \cite{Alexandrou:2022dtc}  & \   $0.924(94)$ & \   $1.051(94)$ \\
\hline\hline
\end{tabular}
\caption{\label{Table:2} The properties $g_T^{u-d}$ and $\kappa_T^{u-d}$ at the scale 
$\mu = 2\,\rm GeV$ from 
(i) the quark models in Table~\ref{Table:1}, and 
(ii) the lattice QCD study of Ref.~\cite{Alexandrou:2022dtc}. 
The model results are evolved in leading order. The lattice results
refer to $\overline{\rm MS}$ scheme. The numbers in brackets represent 
the theoretical uncertainties in the last digits, see text.}
\end{table}

An important test of a model calculation consists in verifying that integrating GPDs over $x$
yields form~factors. For that we carried out a two-step test. In the first step, we checked that 
integrating $H_T^q(x,\xi,t)$, $E_T^q(x,\xi,t)$, $\tilde{H}_T^q(x,\xi,t)$ over $x$
yield form factors, i.e.\ functions of the variable $t$ which are independent of $\xi$. 
In the second step, we verified that the form factors obtained from GPD integration in 
Eq.~(\ref{odd-first-moment}) and evaluation of Eq.~(\ref{Eq:def-tensor-FFs}) agree with 
each other within numerical accuracy which in our calculation was about ${\cal O}(10^{-3})$. 
These are important consistency tests of a model calculation which the bag model 
successfully satisfies.
(In Sec.~\ref{Sec-4:chiral-odd-GPDs-in-bag} we discussed the special case
$\int dx \,\tilde{E}_T^q(x,\xi,t)=0$ \cite{Diehl:2001pm}
where the form factor vanishes.)

Obtaining form factors from the lowest Mellin moments of GPDs is a special case of the
polynomiality property. This is
a general statement that Mellin moments of GPDs are polynomials in $\xi$ \cite{Ji:1998pc} 
with Eq.~(\ref{odd-first-moment}) being a special case for $N=1$ Mellin moments. 
For the $N=2$ Mellin moments of chiral-odd GPDs, one has
\cite{Diehl:2005jf,Hagler:2004yt,Burkardt:2005hp}
\be
\label{odd-second-moment}
  \int dx\,x\,F^q(x,\xi,t)  \ = \ 
  \begin{cases} 
  F_{20}^q(t)   & \mbox{for} \quad F^q = H_T^q,\,E_T^q,\,\tilde{H}_T^q, \cr
  F_{21}^q(t)\;\xi                  & \mbox{for} \quad F^q = \tilde{E}_{T}^q.
  \end{cases}
\ee
The notation reflects that generally $F_{Nk}^q(t)$ is a coefficient in the $N^{\rm th}$ Mellin 
moment $\int dx\,x^{N-1}\,F^q(x,\xi,t)$ associated with the power $\xi^k$ where $k\le N$.
Odd powers $k$ appear only for $F^q = \tilde{E}_T^q$. 
For $F^q = H_T^q,\,E_T^q,\,\tilde{H}_T^q$ and chiral-even twist-2 GPDs,
the powers $k$ are always even. The highest possible even powers $k=N$ (for even $N$)
are assumed only for the chiral-even unpolarized GPDs associated with the structure 
$\Gamma=\gamma^+$ and are related to the $D$-term~\cite{Polyakov:1999gs}.
    (The notation (\ref{odd-second-moment}) would correspond to $H_{T10}^q(t)\equiv H_T^q(t)$,
    $E_{T10}^q(t)\equiv E_{T}^q(t)$, $\tilde{H}_{T10}^q(t)\equiv \tilde{H}_{T}^q(t)$
    in Eqs.~(\ref{odd-first-moment},~\ref{Eq:def-tensor-FFs}) from which we refrained
    for brevity. Sometimes in literature the notation
    $H^q_{TNk}(t)=A^q_{TNk}(t)$, $E^q_{TNk}(t)=B^q_{TNk}(t)$ and analogously for the
    chiral-odd GPDs with ``tilde'' 
    is used. In some works it was convenient to introduce extra prefactors, for instance our
    $\tilde{E}^q_{T21}(t)=-2\tilde{B}^q_{T21}(t)$ in 
    \cite{Diehl:2005jf,Hagler:2004yt,Burkardt:2005hp}.)
  
We checked numerically that our results for the second Mellin moments $N=2$ of chiral-odd GPDs 
comply with Eq.~(\ref{odd-second-moment}), see App.~\ref{App:polynomiality}. 
The bag model has passed analogous tests also for 
chiral-even GPDs \cite{Ji:1997gm,Tezgin:2020vhb} which specifically in the case of the
unpolarized GPDs are related to form factors of the energy-momentum tensor \cite{Neubelt:2019sou,Lorce:2022cle}.
For $N>2$ the Mellin moments are in general divergent in the bag model, and would require to introduce a 
suitable regularization method to evaluate them numerically. We refrain from this step in this work. 
For analytical proofs of polynomiality for all Mellin moments of chiral-even GPDs in the chiral quark 
soliton model, we refer to \cite{Schweitzer:2002yu,Schweitzer:2002nm,Schweitzer:2003ms,Ossmann:2004bp}.

\newpage
\section{Conclusions}
\label{Sec-10:conclusions}

We presented a study of twist-2 chiral-odd GPDs of the proton in the bag model. In a prior result in 
literature, 3 out of the 4 chiral-odd GPDs were incorrectly found to vanish in the bag model 
\cite{Scopetta:2005fg} in contrast to other models. 
In this work we corrected this result and showed that all 4 chiral-odd GPDs are non-zero in the 
bag model.

The bag model results are theoretically consistent, comply with inequalities, and satisfy polynomiality. 
The results for $H_T^q(x,\xi,t)$, $E_T^q(x,\xi,t)$, $\tilde{H}_T^q(x,\xi,t)$ agree within 20-40$\%$ with 
other models in the valence-$x$ region $0.1\lesssim x \lesssim 0.7$ and small to moderate~$t$ 
\cite{Pasquini:2005dk,Pasquini:2007xz,Lorce:2011dv,Chakrabarti:2015ama,Maji:2017ill,Wakamatsu:2008ki,
Kaur:2023lun,Liu:2024umn}.  
A similar model dependence was observed across quark model predictions of other partonic 
properties like transverse momentum dependent PDFs \cite{Boussarie:2016qop}. 
As they mimic the complexities of QCD dynamics in terms of simpler model interactions, 
models have limitations and caveats, e.g., the bag model gives rise to unphysical antiquark PDFs 
and is at variance with chiral symmetry (though ways exist
\cite{Theberge:1980ye,Thomas:1981vc,Thomas:1982kv,Signal:2021aum} to improve upon these aspects).
Despite their limitations and caveats, it is gratifying to observe that different models 
based on a variety of model assumptions give rise to a consistent picture for partonic properties 
of the nucleon at low initial scales.

The GPD $\tilde{E}_T^q(x,\xi,t)$ is particularly challenging to model and we observe a larger
spread of model predictions. In fact, this is the only twist-2 GPD which (i) is odd under 
$\xi\to(-\xi)$ and (ii) whose lowest Mellin moment vanishes. The bag model complies with
both properties, in contrast to many currently available models based on a Fock space expansion 
truncated after the lowest $qqq$-Fock component. In these models, GPDs can only be studied for $x>\xi$ 
and it is not possible to comply with the properties (i) and (ii),
and polynomiality is not satisfied. While this affects all GPDs, it leads to particularly
apparent qualitative and quantitative differences in a case when the lowest Mellin moment 
of a GPD must vanish and a model cannot comply with~it.  
In principle, polynomiality could be restored in these models after resumming
all Fock states.

It is important to stress that the bag model is not based on a Fock expansion 
but on a direct evaluation of quark correlator matrix elements of the type 
$\la p'|\bar{\psi}_q\Gamma\psi_q|p\ra$ and therefore does not experience the limitations 
associated with a Fock space truncation. (The quark correlator is evaluated in the bag model
in terms of the instant-form quark wave functions which should should not to be confused with 
a lighfront Fock space expansion.)
Indeed, this work is one of the few quark model studies where 
the sum rule $\int dx\,\tilde{E}_T^q(x,\xi,t)=0$ is consistently described.

Although the bag model calculation is carried out for the physical number of
colors $N_c=3$ which does not seem~large, we observe a good agreement with large-$N_c$ 
predictions \cite{Schweitzer:2016jmd,Schweitzer:2015zxa}
regarding the flavor dependence of chiral-odd GPDs. 
For instance, in the bag model the $u$- and $d$-flavors have opposite signs for
$H_T^q(x,\xi,t)$ and $\tilde{E}_T^q(x,\xi,t)$, while they have equal signs for the
combination $\overline{E}_T^q(x,\xi,t) = E_T^q(x,\xi,t)+2\,\tilde{H}_T^q(x,\xi,t)$.
These results agree with the trends for the flavor dependencies predicted in the 
large-$N_c$ limit \cite{Schweitzer:2016jmd,Schweitzer:2015zxa}.
Indications from phenomenology support this flavor dependence for 
$H_T^q(x,\xi,t)$ and $\overline{E}_T^q(x,\xi,t)$ \cite{Kubarovsky:2016yaa}.

In order to further test the model, we compared our results to lattice QCD calculations of 
the flavor-nonsinglet chiral odd GPDs \cite{Alexandrou:2021bbo} based on the quasi distribution 
method~\cite{Ji:2013dva} for $\xi=0$ and $t=-0.69\,\rm GeV^2$. 
We evolved the model results and conservatively estimated the uncertainties
associated with the not well known initial scale of the model.
Presently, a quantitative comparison is hindered because the lattice results
correspond to quasi GPDs with $P_z = 1.67\,$GeV the largest available value 
in Ref.~\cite{Alexandrou:2021bbo}. Keeping this in mind, the 
bag model results for $H_T^{u-d}(x,0,t)$, $E_T^{u-d}(x,0,t)$, $\tilde{H}_T^{u-d}(x,0,t)$
exhibit good agreement with the quasi GPD lattice results \cite{Alexandrou:2021bbo}
in the region $0.2 \lesssim x\lesssim 0.5$ where quark models can be expected to work best. 
When in future lattice calculations with larger $P_z$ will become available and controlled 
extrapolations $P_z\to\infty$ feasible, it will be possible to make more quantitative comparisons.

The GPD $\tilde{E}_T^q(x,\xi,t)$ was found in the lattice QCD study at $\xi=1/3$ to be compatible 
with zero within the statistical accuracy of the lattice simulation~\cite{Alexandrou:2021bbo}. 
This finding is also in qualitative agreement with the bag model in the sense that this is by 
far the smallest chiral-odd GPD in the bag model. In addition to that, the bag model predicts 
for this GPD a node in the valence-$x$ region where the other GPDs are largest and give the
clearest signal in a lattice QCD calculation. The bag model results therefore indicate that 
a precise determination of $\tilde{E}_T^q(x,\xi,t)$ via the quasi GPD method might be much
more challenging than that of other chiral-odd GPDs.

We studied also tensor form factors and presented a detailed comparison of two 
properties related to them, namely $g_T^q$ and $\kappa_T^q$. In both cases, the bag model 
is in good agreement with other quark models \cite{Pasquini:2005dk,Ledwig:2010zq,Gutsche:2016xff,Kaur:2023lun}.
After taking evolution effects into account, we also observe a good agreement of the bag and 
other quark models with lattice QCD results for the respective flavor-nonsinglet combinations 
\cite{Alexandrou:2022dtc}. 

To summarize, the bag model gives a theoretically consistent description of chiral-odd GPDs
of the nucleon which agrees well with other models except for the challenging-to-model GPD 
$\tilde{E}_T^q(x,\xi,t)$ where, to the best of our knowledge, we presented the first 
dynamical quark model study describing the polynomiality properties of this GPD consistently. 
The bag model predictions also agree with current state-of-the-art lattice QCD calculations
within model accuracy. 
Our results help to give confidence that quark models can successfully catch the main features 
of chiral-odd GPDs.

\newpage

\noindent {\bf Acknowledgments.}
We thank Martha Constantinou, Barbara Pasquini, Asli Tandogan, 
Valerio Bertone, Wally Melnitchouk and Christian Weiss for helpful discussions. 
The work of K.T.\ was supported by the U.S.\ Department of Energy 
under Contract No.\ DE-SC0012704.
The work of B.M.\ and P.S.\ was supported by the National Science Foundation 
under Awards 2111490 and 2412625, and the U.S. Department of Energy
under the umbrella of the Quark-Gluon Tomography (QGT) 
Topical Collaboration with Award No.\ DE-SC0023646.

\appendix

\section{Model expressions for \boldmath
  ${\cal M}_{\lambdaS^\prime\lambdaS}[i\sigma^{+j}]$}
\label{App:model-expressions-for-matrix-elements}

In the bag model, the evaluation of the nucleon matrix elements  
${\cal M}_{\lambdaS^\prime\lambdaS}[\Gamma]$ defined in Eq.~(\ref{Eq:abbreviation}) 
for chiral-odd twist-2 GPDs requires the determination of the  
${\cal A}_{\lambdaS^\prime\lambdaS}[i\sigma^{+j}]$ defined in 
Eq.~(\ref{GPDmatrixelements}) which yields for $j=1,\,2$ the results

\ba \label{i1}
    {\cal A}_{\lambdaS^\prime\lambdaS}[i\sigma^{+1}] 
    &=& 
    4\pi A^2 R^6 \bigg\{
    \Big( i\sigma^2 t_0(k) t_0(k') + \vec{\sigma}\cdot\hat{k}^\prime\sigma^1   t_0(k) t_1(k') 
    - \sigma^1\vec{\sigma}\cdot\hat{k}   t_0(k') t_1(k) - i \vec{\sigma}\cdot\hat{k}^\prime\sigma^2\vec{\sigma}\cdot\hat{k} t_1(k)t_1(k')\Big)_{s's} \nonumber \\
    &+& 
    \Big[ \frac{\Delta_x}{\sqrt{-t}}\sinh w\Big] \Big(    t_0(k)t_0(k')- \vec{\sigma}\cdot\hat{k}^\prime \vec{\sigma}\cdot\hat{k}   t_1(k) t_1(k')\Big)_{s's} \nonumber \\
    &+& 
    \Big[ \frac{i\Delta_y}{\sqrt{-t}}\sinh w\Big] \Big(    \vec{\sigma}\cdot\hat{k}^\prime   t_0(k) t_1(k') - \vec{\sigma}\cdot\hat{k}   t_0(k') t_1(k)\Big)_{s's} \nonumber \\
    &+& 
    \Big[ \frac{2\Delta_x}{t}\sinh^2\frac{w}{2}\Big] \Big(- \vec{\sigma}\cdot\hat{k}^\prime \vec{\sigma}\cdot\Delta \, t_0(k) t_1(k')
    + \vec{\sigma}\cdot\Delta \vec{\sigma}\cdot\hat{k} \, t_0(k') t_1(k)\Big)_{s's} \nonumber \\
    &+& 
    \Big[ \frac{2i\Delta_y}{t}\sinh^2\frac{w}{2}\Big] \Big(- \vec{\sigma}\cdot\Delta \, t_0(k) t_0(k')
    + \vec{\sigma}\cdot\hat{k}^\prime \vec{\sigma}\cdot\Delta \vec{\sigma}\cdot\hat{k} \, t_1(k) t_1(k')\Big)_{s's} \bigg\}\, , 
    \\  { } \nonumber\\
    \label{i2}
    {\cal A}_{\lambdaS^\prime\lambdaS}[i\sigma^{+2}] 
    &=& 
    4\pi A^2 R^6 
    \bigg\{\Big( -i\sigma^1 t_0(k) t_0(k') + \vec{\sigma}\cdot\hat{k}^\prime\sigma^2   t_0(k) t_1(k') -
     \sigma^2\vec{\sigma}\cdot\hat{k} t_0(k') t_1(k) + i \vec{\sigma}\cdot\hat{k}^\prime\sigma^1\vec{\sigma}\cdot\hat{k} t_1(k)t_1(k')\Big)_{s's} \nonumber \\
    &+& 
    \Big[ \frac{\Delta_y}{\sqrt{-t}}\sinh w\Big] \Big(    t_0(k)t_0(k')- \vec{\sigma}\cdot\hat{k}^\prime \vec{\sigma}\cdot\hat{k}   t_1(k) t_1(k')\Big)_{s's} \nonumber \\
    &+& 
    \Big[ \frac{i\Delta_x}{\sqrt{-t}}\sinh w\Big] \Big( -\vec{\sigma}\cdot\hat{k}^\prime   t_0(k) t_1(k') + \vec{\sigma}\cdot\hat{k}   t_0(k') t_1(k)\Big)_{s's} \nonumber \\
    &+& 
    \Big[ \frac{2\Delta_y}{t}\sinh^2\frac{w}{2}\Big] \Big(- \vec{\sigma}\cdot\hat{k}^\prime \vec{\sigma}\cdot\Delta \, t_0(k) t_1(k')
    + \vec{\sigma}\cdot\Delta \vec{\sigma}\cdot\hat{k} \, t_0(k') t_1(k)\Big)_{s's} \nonumber \\
    &+& 
    \Big[ \frac{2i\Delta_x}{t}\sinh^2\frac{w}{2}\Big] \Big(\vec{\sigma}\cdot\Delta \, t_0(k) t_0(k')
    - \vec{\sigma}\cdot\hat{k}^\prime \vec{\sigma}\cdot\Delta \vec{\sigma}\cdot\hat{k} \, t_1(k) t_1(k')\Big)_{s's} \bigg\}\, , 
\ea
where we defined $\chi^\dag_{s'}(\dots)\chi^{ }_{s^{ }} = (\dots)_{s's}$. 

\newpage
\section{Expressions for the \boldmath $T_i^q$ terms}
\label{App:Tq}

Based on the results for ${\cal A}_{\lambdaS^\prime\lambdaS}[i\sigma^{+j}]$ 
in Eqs.~(\ref{i1},~\ref{i2}) we obtain after projection on Pauli matrices
$\sigma^0$, $\sigma^1$, $\sigma^2$, $\sigma^3$ (where $\sigma^0$ denotes the
$2\times2$ unit matrix) and taking traces over spin indices the following results
\ba\label{Eq:defineTi}
T_1^q(\vec{\Delta}) = \frac{1}{4m} \; {\rm tr}_{\rm spin} \Bigl[ {\cal M}[i\sigma^{+1}]\,\sigma^0 \Bigr] \, , &&
T_2^q(\vec{\Delta}) = \frac{1}{4 i m} \; {\rm tr}_{\rm spin} \Bigl[ {\cal M}[i\sigma^{+1}]\,\sigma^2 \Bigr] \, , \nonumber\\
T_3^q(\vec{\Delta}) = \frac{1}{4 i m} \: {\rm tr}_{\rm spin} \Bigl[ {\cal M}[i\sigma^{+2}]\,\sigma^1 \Bigr] \, , &&
T_4^q(\vec{\Delta}) = \frac{1}{4 i m} \; {\rm tr}_{\rm spin} \Bigl[ {\cal M}[i\sigma^{+2}]\,\sigma^3 \Bigr] \, ,
\ea
whereby other combinations of $j=1,\,2$ and $\sigma^k$ give no new information. The $T_i^q(\vec{\Delta})$ are given by
\ba
\label{integralT1}
    T_1^q(\vec{\Delta}) 
    &=& 
    Z^2(t)\frac{4\pi A^2 R^6 \bar{m} N_q}{1-(\cosh w-1)\Delta_z^2/t}\int\frac{d^2k_\perp}{(2\pi)^3}\bigg\{ t_0(k)t_0(k') \Big[ \frac{\Delta_x}{\sqrt{-t}}\sinh w \Big] \nonumber \\
    &+& \frac{t_0(k)t_1(k')}{k'}\Big[ k_{x}' - 2 \frac{\vec{k}^\prime\cdot\vec{\Delta}}{t}\Delta_x\sinh^2\frac{w}{2}\Big] \nonumber \\
    &-& \frac{t_0(k')t_1(k)}{k}\Big[ k_{x} - 2 \frac{\vec{k}\cdot\vec{\Delta}}{t}\Delta_x\sinh^2\frac{w}{2}\Big] \nonumber \\
    &-& \frac{t_1(k)t_1(k')}{kk'}\Big[ (\vec{k}^\prime\times\vec{k})_y + \vec{k}\cdot\vec{k}^\prime\frac{\Delta_x}{\sqrt{-t}}\sinh w +  
        2\,\frac{\vec{k}^\prime\cdot(\vec{\Delta}\times\vec{k})}{t}\Delta_y \sinh^2\frac{w}{2} \Big]\bigg\}\, ,
        \\
\label{integralT2}
    T_2^q(\vec{\Delta}) 
    &=& 
    Z^2(t)\frac{4\pi A^2 R^6 \bar{m} P_q}{1-(\cosh w-1)\Delta_z^2/t}\int\frac{d^2k_\perp}{(2\pi)^3}
    \bigg\{ t_0(k)t_0(k')\Big[ 1 - 2\frac{\Delta_y^2}{t}\sinh^2\frac{w}{2}\Big] \nonumber \\
    &+& \frac{t_0(k)t_1(k')}{k'}\Big[ k_{z}' + k_{y}' \frac{\Delta_y}{\sqrt{-t}}\sinh w 
        - 2\, \frac{(\vec{k}^\prime\times\vec{\Delta})_y}{t}\Delta_x\sinh^2\frac{w}{2} \Big] \nonumber \\
    &+& \frac{t_0(k')t_1(k)}{k}\Big[ k_{z} - k_{y} \frac{\Delta_y}{\sqrt{-t}}\sinh w 
        - 2\, \frac{(\vec{k}\times\vec{\Delta})_y}{t}\Delta_x\sinh^2\frac{w}{2}\Big] \nonumber \\
    &+& \frac{t_1(k)t_1(k')}{kk'}\Big[ \vec{k}\cdot\vec{k}^\prime-2k_y k_y' 
        + \frac{(\vec{k}\times\vec{k}^\prime)_y}{\sqrt{-t}}\Delta_x\sinh w 
        + 2 \frac{\Delta_y}{t} (\vec{k}^\prime\cdot\vec{\Delta}\,k_y-\vec{k}\cdot\vec{k}^\prime \,\Delta_y 
        + \vec{\Delta}\cdot\vec{k}\,k^\prime_y)\sinh^2\frac{w}{2}\Big]\bigg\}\, , \quad 
        \\
\label{integralT3}
    T_3^q(\vec{\Delta}) 
    &=& Z^2(t)\frac{4\pi A^2 R^6 \bar{m} P_q}{1-(\cosh w-1)\Delta_z^2/t}\int\frac{d^2k_\perp}{(2\pi)^3} 
        \bigg\{ -t_0(k)t_0(k')\Big[ 1 - 2\frac{\Delta_x^2}{t}\sinh^2\frac{w}{2}\Big] \nonumber \\
    &-& \frac{t_0(k)t_1(k')}{k'}\Big[ k_{z}' + k_{x}^\prime \frac{\Delta_x}{\sqrt{-t}}\sinh w 
        + 2\, \frac{(\vec{k}^\prime\times\vec{\Delta})_x}{t}\Delta_y\sinh^2\frac{w}{2} \Big] \nonumber \\
    &-& \frac{t_0(k')t_1(k)}{k}\Big[ k_{z} - k_{x} \frac{\Delta_x}{\sqrt{-t}}\sinh w 
        + 2\, \frac{(\vec{k}\times\vec{\Delta})_x}{t}\Delta_y\sinh^2\frac{w}{2}\Big] \nonumber \\
    &-& \frac{t_1(k)t_1(k')}{kk'}\Big[ \vec{k}\cdot\vec{k}^\prime-2k_x k_x^\prime 
        - \frac{(\vec{k}\times\vec{k}^\prime)_x}{\sqrt{-t}}\Delta_y\sinh w 
        + 2 \frac{\Delta_x}{t} (\vec{k}^\prime\cdot\vec{\Delta}\,k_x-\vec{k}\cdot\vec{k}^\prime \,\Delta_x 
        + \vec{\Delta}\cdot\vec{k}\,k^\prime_x)\sinh^2\frac{w}{2}\Big]\bigg\}\, , \\
\label{integralT4}
    T_4^q(\vec{\Delta}) 
    &=& Z^2(t)\frac{4\pi A^2 R^6 \bar{m} P_q}{1-(\cosh w-1)\Delta_z^2/t}\int\frac{d^2k_\perp}{(2\pi)^3}
        \bigg\{ 2\,t_0(k)t_0(k')\Big[ \frac{\Delta_x\Delta_z}{t}\sinh^2\frac{w}{2} \Big] \nonumber \\
    &+& \frac{t_0(k)t_1(k')}{k'}\Big[ k_{x}^\prime - k_z^\prime \frac{\Delta_x}{\sqrt{-t}}\sinh w 
        - 2 \frac{(\vec{k}^\prime\times\Delta)_z}{t}\Delta_y\sinh^2\frac{w}{2}\Big] \nonumber \\
    &+& \frac{t_0(k')t_1(k)}{k}\Big[ k_{x} + k_z \frac{\Delta_x}{\sqrt{-t}}\sinh w 
        - 2 \frac{(\vec{k}\times\Delta)_z}{t}\Delta_y\sinh^2\frac{w}{2}\Big] \nonumber \\
    &+& \frac{t_1(k)t_1(k')}{kk'}\Big[k_x k_z^\prime + k_x^\prime k_z 
        + \frac{(\vec{k}\times\vec{k}^\prime)_z}{\sqrt{-t}}\Delta_y\sinh w 
        - 2\frac{\Delta_x}{t} (\vec{k}^\prime\cdot\vec{\Delta}\,k_z -\vec{k}\cdot\vec{k}^\prime \,\Delta_z 
        + \vec{\Delta}\cdot\vec{k}\,k^\prime_z)\sinh^2\frac{w}{2}\Big]\bigg\}\, .
\ea
In the above expressions the $\delta$-function in Eq.~(\ref{GPDmatrixelements}) is 
integrated out for notational simplicity, i.e.\ $\vec{k}=(k_x,k_y,K)$ with $K$ defined in 
Eq.~(\ref{GPDmatrixelements3}), and $d^2k_\perp = dk_x\,dk_y=d\varphi\,k_\perp dk_\perp$.
$Z(t)$ encodes the effects of spectator quarks, and is given by
\begin{equation}
    \label{Eq:Z-of-t}
    Z(t)=\frac{A^2}{\cosh w}\int_{0}^{R} dr\,r^2\,j_0(\varepsilon_{0}|\vec{\Delta}|r/\bar{m})\,
    (j_0^2(\varepsilon_{0} r)+j_1^2(\varepsilon_{0} r))\,.
\end{equation}

\section{Nucleon spinor expressions}
\label{App:nucleon-spinor-expressions}

In order to evaluate the nucleon spinor expressions on the right-hand-side of
Eq.~(\ref{oddGPDs}), we introduce the notation
\sub{\label{Eqs:coeff-chiral-odd}
\ba
    {\cal S}^{+i}_{\lambdaS^\prime\lambdaS}[H_T^q] &=& \bar u(p',\lambdaS')\,
    i\sigma^{+i}\,u(p,\lambdaS) \, , \\
    {\cal S}^{+i}_{\lambdaS^\prime\lambdaS}[\tilde H_T^q] &=& \bar u(p',\lambdaS')\,
    \frac{P^{+}\Delta^i+ \Delta^{+}P^i}{m^2}\,u(p,\lambdaS) \, , \\
    {\cal S}^{+i}_{\lambdaS^\prime\lambdaS}[E_T^q] &=& \bar u(p',\lambdaS')\,
    \frac{\gamma^{+}\Delta^i-\Delta^{+}\gamma^i}{2m}\,u(p,\lambdaS) \, , \\
    {\cal S}^{+i}_{\lambdaS^\prime\lambdaS}[\tilde E_T^q] &=& \bar u(p',\lambdaS')\,
    \frac{\gamma^+ P^i- P^+\gamma^i}{m}\,u(p,\lambdaS) \, .
\ea}
We use the standard nucleon spinor expressions normalized as 
$\bar{u}(p,s')\,u(p,s)=2m\,\delta_{ss'}$ given by
\be
    u(p,s) = \frac{\fslash{p}+m}{\sqrt{p^0+m}}\,
    \left(\begin{matrix} \chi_s \\ 0 \end{matrix}\right)\,
\ee
where $\chi^\dag_{s'}\chi^{ }_{s^{ }} = \delta_{ss'}$ and we use
$\chi^\dag_{s'}(\dots)\chi^{ }_{s^{ }} = (\dots)_{s's}$ for brevity
as defined in App.~\ref{App:model-expressions-for-matrix-elements}.
In this notation, the expressions for the nucleon spinor structures on the 
right-hand side of Eq.~(\ref{oddGPDs}) are given for the
transverse index $j=1$ by
\begin{alignat}{3}
    {\cal S}^{+1}_{\lambdaS^\prime\lambdaS}[H_T^q] 
    &= &&
    (\bar{m}+m)\, \Big(
      i\sigma^2
    - i\frac{\vec{\sigma}\cdot\vec{p}\,^{\prime}}{\bar{m}+m}\sigma^2
      \frac{\vec{\sigma}\cdot\vec{p}}{\bar{m}+m}
    + \frac{\vec{\sigma}\cdot\vec{p}\,^{\prime}}{\bar{m}+m}\sigma^1
    - \sigma^1\frac{\vec{\sigma}\cdot\vec{p}}{\bar{m}+m}
    \Big)_{s's} \, , 
    \\
    {\cal S}^{+1}_{\lambdaS^\prime\lambdaS}[\tilde H_T^q] 
    &=&       
      \frac{\bar m\Delta_x}{m^2}  
    & (\bar m + m)\, \Big(
      \sigma^0
    -  \frac{\vec{\sigma}\cdot\vec{p}\,^{\prime}}{\bar{m}+m}
      \frac{\vec{\sigma}\cdot\vec{p}}{\bar{m}+m}\Big)_{s's} 
      \, ,\\
    {\cal S}^{+1}_{\lambdaS^\prime\lambdaS}[E_T^q] 
    &=& 
      \frac{\Delta_x}{2m} 
    & (\bar m+m)\, \Big(
    \sigma^0
    + \frac{\vec{\sigma}\cdot\vec{p}\,^{\prime}}{\bar{m}+m}\sigma^3 
    + \sigma^3\frac{\vec{\sigma}\cdot\vec{p}}{\bar{m}+m} 
    + \frac{\vec{\sigma}\cdot\vec{p}\,^{\prime}}{\bar{m}+m}\,
      \frac{\vec{\sigma}\cdot\vec{p}}{\bar{m}+m}
    \Big)_{s's}  \, ,
 \nonumber\\
    &&- \frac{\Delta_z}{2m} &
    (\bar m+m)\, \Big(
      \frac{\vec{\sigma}\cdot\vec{p}\,^{\prime}}{\bar{m}+m}\sigma^1
      + \sigma^1\frac{\vec{\sigma}\cdot\vec{p}}{\bar{m}+m}
      \Big)_{s's}  \,, \\
    {\cal S}^{+1}_{\lambdaS^\prime\lambdaS}[\tilde E_T^q] 
    &=& - \frac{\bar m}{m}\, &
        (\bar m+m)\, \Big(
        \frac{\vec{\sigma}\cdot\vec{p}\,^{\prime}}{\bar{m}+m}\sigma^1
      + \sigma^1\frac{\vec{\sigma}\cdot\vec{p}}{\bar{m}+m}\Big)_{s's} \,.
\end{alignat}
For the transverse index $i=2$, the corresponding expressions are given by
\begin{alignat}{3}
    {\cal S}^{+2}_{\lambdaS^\prime\lambdaS}[H_T^q] 
    &=& & (\bar{m}+m) \Big(
    - i\sigma^1 
    + \frac{\vec{\sigma}\cdot\vec{p}\,^{\prime}}{\bar{m}+m}\sigma^2
    - \sigma^2\frac{\vec{\sigma}\cdot\vec{p}} {\bar{m}+m}
    + i\frac{\vec{\sigma}\cdot\vec{p}\,^{\prime}} 
    {\bar{m}+m}\sigma^1\frac{\vec{\sigma}\cdot\vec{p}}{\bar{m}+m}\Big)_{s's}
     \, ,
    \\
    {\cal S}^{+2}_{\lambdaS^\prime\lambdaS}[\tilde H_T^q] 
    &=& \frac{(\bar m\Delta_y)}{m^2} &
    (\bar m + m)\, \Big(\sigma^0-\frac{\vec{\sigma}\cdot\vec{p}\,^{\prime}}{\bar{m}+m}\frac{\vec{\sigma}\cdot\vec{p}}{\bar{m}+m}\Big)_{s's}
    \,,\\
    {\cal S}^{+2}_{\lambdaS^\prime\lambdaS}[E_T^q] 
    &=& \frac{\Delta_y}{2m} &
    (\bar m+m)\, \Big(
      \sigma^0 
    + \frac{\vec{\sigma}\cdot\vec{p}\,^{\prime}}{\bar{m}+m}\sigma^3
    + \sigma^3\frac{\vec{\sigma}\cdot\vec{p}}{\bar{m}+m}
    + \frac{\vec{\sigma}\cdot\vec{p}\,^{\prime}}{\bar{m}+m}
      \frac{\vec{\sigma}\cdot\vec{p}}{\bar{m}+m} 
      \Big)_{s's}      
    \nonumber \\
    &&- \frac{\Delta_z}{2m} &
    (\bar m+m)\, \Big(
      \frac{\vec{\sigma}\cdot\vec{p}\,^{\prime}}{\bar{m}+m}\sigma^2
    + \sigma^2\frac{\vec{\sigma}\cdot\vec{p}}{\bar{m}+m}\Big)_{s's}    
    \,,\\
    {\cal S}^{+2}_{\lambdaS^\prime\lambdaS}[\tilde E_T^q] 
    &=& - \frac{\bar m}{m} &
    (\bar m+m)\,  \,\Big(
      \frac{\vec{\sigma}\cdot\vec{p}\,^{\prime}}{\bar{m}+m}\sigma^2
    + \sigma^2\frac{\vec{\sigma}\cdot\vec{p}}{\bar{m}+m}\Big)_{s's}  
     \,,
\end{alignat}
where $\sigma^0$ denotes the unit $2\times2$ matrix.
These expressions are valid in a general frame. Using the Breit frame 
introduced in Eq.~(\ref{Eq:Breit-frame}), projecting and taking traces 
with the respective $\sigma^k_{s's}$-matrices as described in App.~\ref{App:Tq}, and then by dividing the overall results by a factor of $2m$, 
one obtains the prefactors in front of the GPDs quoted on the right-hand sides of 
Eqs.~(\ref{T1}--\ref{T4}).

\newpage
\section{\boldmath Bag model results for tensor form factors}
\label{App:form-factors}

In order to evaluate the tensor form factors, it is convenient to introduce
the notation ($\sigma^0$ is the unit $2\times2$ matrix)
\be\label{Eq:new-notation-Rijk}
    R^{\mu\nu,k}(\vec{\Delta}) = \frac{1}{2i}\,{\rm tr}_{\rm spin} \biggl[
    \la p^\prime,\vec s^{\,\prime}|  \bar{\psi}(0) i\sigma^{\mu\nu}\psi(0)|p,\vec s\rangle
    \,\sigma^k_{ss^\prime} \biggr] \,.
\ee 
In this notation, the three linear equations for the three tensor form factors 
can be expressed as
\begin{align}
    \frac{R^{12,3}(\vec{\Delta})}{2}
    & = 
        \bigg(m + \frac{\Delta_z^2}{4(\bar{m}+m)}\bigg)\,H_T^q(t) + \frac{t+\Delta_z^2}{4\,m} \, E_T^q(t)
        \nonumber\\
    &   = a(t) P_q \int\frac{d^3k}{(2\pi)^3}\bigg\{
        \Big[t_0(k)t_0(k^\prime)-\frac{t_1(k)t_1(k^\prime)}{k\,k^\prime}(2\,k^\prime_z\,k_z - \vec{k}^\prime\cdot \vec{k})\Big] 
        \Big( 1 + \frac{2 \Delta_z^2}{|\Delta|^2} \sinh^2 \frac{w}{2}\Big) \nonumber \\ 
    &   \hspace{2.8cm} - \frac{2\,t_1(k)t_1(k')}{k\,k^\prime}\Big[ (k_x^\prime k_z + k_z^\prime k_x)\frac{\Delta_x\Delta_z}{|\Delta|^2} 
        + (k_y^\prime k_z + k_z^\prime k_y)\frac{\Delta_y\Delta_z}{|\Delta|^2} \Big] \sinh^2 \frac{w}{2}  \nonumber \\
    &   \hspace{2.8cm} + \Big[ \frac{t_0(k)t_1(k')}{k^\prime}\,k^\prime_z-\frac{t_0(k^\prime)t_1(k)}{k}\,k_z\Big] 
        \frac{\Delta_z}{|\Delta|}\sinh w\bigg\} \, ,  
        \label{Eq:FF1} \\
        \nonumber\\
    \frac{R^{31,2}(\vec{\Delta})}{2}
    & = 
        - \bigg(m + \frac{\Delta_y^2}{4\,(\bar{m} + m)} \bigg)\,H_T^q(t) - \frac{t + \Delta_y^2}{4\,m} \, E_T^q(t) 
        \nonumber\\
    & = a(t) P_q \int \frac{d^3k}{(2\pi)^3}\bigg\{ \Big[-t_0(k)t_0(k^\prime)+\frac{t_1(k)t_1(k^\prime)}{k\,k^\prime}
        (2\,k_y^\prime\,k_y - \vec{k}^\prime\cdot \vec{k})\Big] \Big(1 + \frac{2\Delta_y^2}{|\Delta|^2}\sinh^2 \frac{w}{2} \Big) \nonumber \\ 
    &   \hspace{2.8cm} + \frac{2\,t_1(k)t_1(k')}{k\,k^\prime}\Big[ (k_x^\prime k_y + k_y^\prime k_x)\frac{\Delta_x\Delta_y}{|\Delta|^2} 
        + (k_z^\prime k_y + k_y^\prime k_z)\frac{\Delta_y\Delta_z}{|\Delta|^2} \Big] \sinh^2 \frac{w}{2}  \nonumber \\
    &   \hspace{2.8cm} + \Big[ \frac{t_0(k^\prime)t_1(k)}{k}\,k_y -\frac{t_0(k)t_1(k')}{k^\prime}\,k^\prime_y\Big] \frac{\Delta_y}{|\Delta|}\sinh w       \bigg\} \, , 
    \label{Eq:FF2}    \\
        \nonumber\\
    \frac{R^{30,0}(\vec{\Delta})}{2i}
    & =
    \frac{\Delta_z}{2}\,H_T^q(t) + \frac{\bar{m}^2\Delta_z}{m^2}\,\tilde{H}_T^q(t) + \frac{\Delta_z}{2}\,E_T^q(t) \nonumber\\
    & = a(t) N_q \int \frac{d^3k}{(2\pi)^3}\bigg\{ \Big[\frac{t_0(k)t_1(k')}{k^\prime}\,k^\prime_z
        -\frac{t_0(k^\prime)t_1(k)}{k}\,k_z\Big] \Big( 1 + \frac{2 \Delta_z^2}{|\Delta|^2} \sinh^2 \frac{w}{2}\Big) \nonumber \\ 
    & \hspace{2.8cm} + \frac{2\,t_0(k)t_1(k')}{k^\prime}\frac{\Delta_z\,(k_\perp^\prime\cdot\Delta_\perp)}{|\Delta|^2} \sinh^2 \frac{w}{2}  
      - \frac{2\,t_0(k')t_1(k)}{k}\frac{\Delta_z\,(k_\perp\cdot\Delta_\perp)}{|\Delta|^2} \sinh^2 \frac{w}{2}  \nonumber \\
    & \hspace{2.8cm} + \Big[t_0(k)t_0(k^\prime)-\frac{t_1(k)t_1(k^\prime)}{k\,k^\prime}(\vec{k}^\prime\cdot \vec{k})\Big] \frac{\Delta_z}{|\Delta|}\sinh w\bigg\} \, .
    \label{Eq:FF3}
\end{align}
where $a(t) = Z^2(t)\,4\,\pi\,m A^2 R^6$ with $Z(t)$ defined  
in App.~\ref{App:Tq} in Eq.~(\ref{Eq:Z-of-t}).

The system of linear equations in Eqs.~(\ref{Eq:FF1}-\ref{Eq:FF3})
becomes numerically unstable for $(-t)<0.05\,{\rm GeV}^2$ and it is
necessary to take the limit $\vec{\Delta}\to 0$ analytically. The
analytical expressions for the form factors at zero-momentum transfer 
are obtained by taking the limits
\ba
&&   \hspace{-5mm}
    \lim\limits_{\vec{\Delta}\to 0\;\;} 
     \biggl[\frac{R^{12,3}(\vec{\Delta})}{2m}\biggr]  
     = 
    H_T^q(0) 
     = P_q\,c_0\int\frac{d^3k}{(2\pi)^3}
      \Bigg[t_0^2+\frac{t_1^2}{3}\Bigg] \, , \\
 &&  \hspace{-5mm} \lim\limits_{\vec{\Delta}\to 0\;\;} 
     \biggl[2m\,\frac{R^{12,3}(\vec{\Delta})+R^{31,2}(\vec{\Delta})}{\Delta_z^2-\Delta_y^2}\biggr] 
     = \frac{H_T^q(0)}{2} + E_T^q(0) 
     = \frac{1}{3} P_q m^2 \eta^2\,c_0 \nonumber\\
 &&  \hspace{18.5mm}
      \times \int\frac{d^3k}{(2\pi)^3}
      \Bigg[\frac{3 t_0^2}{2\,m^2\,\eta^2} 
          + \frac{4 t_0t_1}{k\,m\,\eta}
          + \frac{t_1^2}{2\,m^2\,\eta^2} + \frac{16\,t_1^2}{5\,k^2}
          -\frac{2\,t_0^{\prime}t_1}{m\,\eta}
          +\frac{2\,t_0 t_1^{\prime}}{m\,\eta} 
          -\frac{16 t_1 t_1^\prime}{5k}
          -\frac{8 t_1 t_1^{\prime\prime}}{5}\Bigg] \, , \;\;\;\;\nonumber\\          
&&   \hspace{-5mm} \lim\limits_{\vec{\Delta}\to 0\;\;} 
     \biggl[\frac{R^{30,0}(\vec{\Delta})}{2i\,\Delta_z}\biggr] 
     = \frac{H_T^q(0)}{2} + \tilde{H}_T^q(0) + \frac{E_T^q(0)}{2} 
     = N_q m \,\eta\,c_0 \int\frac{d^3k}{(2\pi)^3}
      \Bigg[\frac{2 t_0 t_1}{3k} + \frac{t_0 t_1^\prime}{3} - \frac{t_0^\prime t_1}{3} + \frac{t_0^2 - t_1^2}{2\,m\,\eta} \Bigg]\, ,
\ea
where $c_0=(4\pi A^2 R^6)$. Solving these equations, we obtain 
\ba
    H_T^q(0) &=& c_0 \int\frac{d^3k}{(2\pi)^3}\;P_q
    \biggl[t_0^2+\frac{t_1^2}{3}\biggr] \, ,   \label{Eq:tensor-charge}\\
    E_T^q(0) &=& 
    c_0 \int\frac{d^3k}{(2\pi)^3}\;P_q \biggl[ 
    \frac{8}{15}\,m^2\eta^2\,\biggl(\frac{2}{k^2}\,t_1^2 + t_1^{\prime\,2}\biggr)
    -\frac{4}{3}\,m\,\eta \,t_0'\,t_1 \biggr]\, , \\
    %
    \tilde{H}_T^q(0) &=& c_0 \int\frac{d^3k}{(2\pi)^3}
    \Big[
    (N_q-P_q)\biggl(\frac{1}{2}\,t_0^2 - \frac{2}{3} \, \eta \,  m \, t_0'\,t_1^{ }\biggr)
    - \left(\frac{N_q}{2}+\frac{P_q}{6}+\frac{8 \eta ^2 m^2 P_q}{15 k^2}\right)\,t_1^2
    -\frac{4}{15} \eta ^2 m^2 P_q t_1'^2
    \Big]
\ea
where we do not indicate the arguments of $t_0(k)$ and $t_1(k)$
and the primes denote derivatives with respect to $k$. The result in 
Eq.~(\ref{Eq:tensor-charge}) corresponds to the tensor charge of the 
nucleon and was computed first in Ref.~\cite{Jaffe:1991kp}.

Interestingly, for GPDs the numerics in the limits $\xi\to0$ and 
$t\to 0$ is so stable that it was not necessary to take these limits
analytically. Indeed, in Fig.~\ref{Fig1:ChiralOddForward} we actually 
plotted the GPDs for $\xi = 10^{-4}$ and $t = 10^{-4}\,{\rm GeV}^2$.
Choosing smaller values like $\xi = 10^{-5}$ and $t = 10^{-5}\,{\rm GeV}^2$
yields the same results within the numerical accuracy which is of the order 
${\cal O}(10^{-4})$ which means that in this way the limit $\xi\to0$ and
$t\to0$ is safely and reliably reached numerically. 
(Recall that $\tilde{E}^q(x,\xi,t)$ vanishes for $\xi=0$ such that it
was fixed $\xi=0.5$ in Fig.~\ref{Fig1:ChiralOddForward}d.)
The numerical computation of form factors for small values of $(-t)$
is more challenging than for GPDs because the form factors effectively 
contain an additional numerical integration.

\section{\boldmath Polynomiality of the second Mellin moments}
\label{App:polynomiality}

Polynomiality for $N=1$ Mellin moments was discussed in detail
in the main text and is satisfied in the bag model. In this Appendix, 
we illustrate how the model satisfies polynomiality for $N=2$ Mellin moments.  

\begin{figure}[t!]
\begin{centering}
\includegraphics[width=8cm]{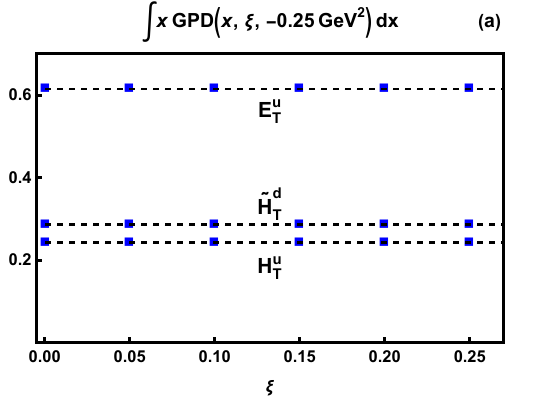} \
\includegraphics[width=8.9cm]{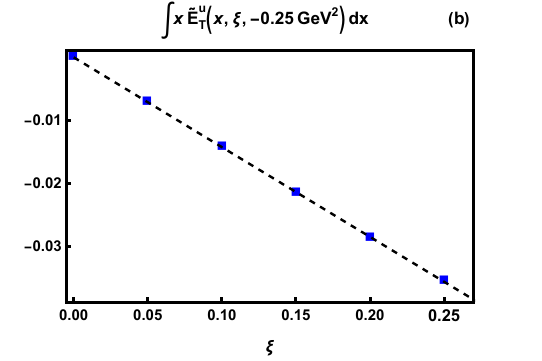}
\par\end{centering}
\caption{\label{Fig9:second-Mellin-moments}
The second Mellin moments of chiral-odd GPDs as functions of $\xi$ at fixed 
$t=-0.25 \text{GeV}^2$ for the indicated flavors. 
The squares are numerical results. The lines are added to guide the eye.
(a) For $H_T^q(x,\xi,t)$, $E_T^q(x,\xi,t)$, $\tilde{H}_T^q(x,\xi,t)$ 
the second Mellin must be constants for fixed $t$ which is the case.
(b) For $\tilde{E}_T^q(x,\xi,t)$ the second Mellin moment must be proportional 
to $\xi$ for fixed $t$ which is also the case. This example illustrates how 
polynomiality condition in Eq.~(\ref{odd-second-moment}) is satisfied in the
bag model. }
\end{figure}

Let us first discuss the polynomiality property in Eq.~(\ref{odd-second-moment}) 
for $H_T^q(x,\xi,t)$, $E_T^q(x,\xi,t)$, $\tilde{H}_T^q(x,\xi,t)$. For these GPDs, the
second Mellin moments must be just form factors, i.e.\ must exhibit no $\xi$-dependence.
The bag model complies with this property. We illustrate this in 
Fig.~\ref{Fig9:second-Mellin-moments}a for a chosen value of $t=-0.25\,{\rm GeV}^2$
where we plot the integrals $\int dx\,x H_T^u(x,\xi,t)$ and similarly for 
$E_T^u(x,\xi,t)$ and $\tilde{H}_T^d(x,\xi,t)$ where the $x$-integrations include 
all $x$,  see footnote~\ref{footnote-unphysical-qbar}. These integrals are even functions
for $\xi$ due Eq.~(\ref{skewness1}) which holds in the bag model as shown
in Sec.~\ref{Sec-4:chiral-odd-GPDs-in-bag} such that we can focus on $\xi\ge 0$. 
The integrals are evaluated numerically for $0\le\xi\le 0.25$ in increments of $0.05$ 
and the results are displayed as squares in Fig.~\ref{Fig9:second-Mellin-moments}a.
Clearly, we observe constant values which correspond to the respective form factors 
$H^u_{T20}(t)$, $E^u_{T20}(t)$, $\tilde{H}^d_{T20}(t)$ at $t=-0.25\,\rm GeV^2$. 
For completeness we remark that the second Mellin of $\tilde{H}_T^u(x,\xi,t)$ happens to 
be nearly zero at the chosen $t$ making in this case the convergence of the numerical 
integrations slower. We have therefore chosen in Fig.~\ref{Fig9:second-Mellin-moments}a 
to display the $d$-flavor for this GPD where the numerics is more accurate.

Next let us turn our attention to the polynomiality property in Eq.~(\ref{odd-second-moment}) 
for $\tilde{E}_T^q(x,\xi,t)$ in which case the second Mellin moments is given by 
$\xi$ times a form factor, i.e.\ must exhibit a linear $\xi$-dependence. As we 
illustrate in Fig.~\ref{Fig9:second-Mellin-moments}a for fixed $t=-0.25\,{\rm GeV}^2$
the bag model complies also with this requirement. 
The integral $\int dx\,x \tilde{E}_T^u(x,\xi,t)$ is an odd function of $\xi$ 
according to Eq.~(\ref{skewness2}) which was shown to be satisfied in the bag model 
in Sec.~\ref{Sec-4:chiral-odd-GPDs-in-bag}, so it is again sufficient to focus on 
$\xi\ge 0$. The results of the numerical calculation are plotted as squares in 
Fig.~\ref{Fig9:second-Mellin-moments}b and show the expected linear behavior.
The slope of the linear function in Fig.~\ref{Fig9:second-Mellin-moments}b 
corresponds to the form factor $\tilde{E}^u_{T21}(t)$ at $t=-0.25\,\rm GeV^2$
[cf.\ the explanation of the naming scheme for form factors of higher Mellin 
moments following Eq.~(\ref{odd-second-moment})].

We verified the polynomiality also at other $t$-values. By systematically repeating 
this procedure one could determine the $t$-behavior of the form factors 
$H^q_{T20}(t)$, $E^q_{T20}(t)$, $\tilde{H}^q_{T20}(t)$, $\tilde{E}^q_{T21}(t)$
in this way, from which we refrain in this work.
We recall that Mellin moments $N\ge 3$ in general diverge in the bag model
and we cannot discuss them.

The numerical results in Fig.~\ref{Fig9:second-Mellin-moments}
refer to $\eta=0.55$ but polynomiality holds for any value of
$\eta$. 
This may at first appear surprising considering the ad hoc way
in which $\eta$ was introduced in Eq.~(\ref{Eq:tildeDelta})
which does not seem compatible with Poincar\'e symmetry.
Polynomiality nevertheless holds for any $\eta$. 
This can be understood from Eq.~(\ref{Eq:tildeDelta}). 
Varying $\eta$ merely rescales the ``units'' in which 
$\xi$ and $t$ are ``measured''. Let us recall that $\eta$
was introduced as an effective way to redistribute the
momentum transfer $\Delta^\mu$ to the nucleon among all
its constituents, see the end of Sec.~\ref{Sec-3:bag-model},
but this step does not spoil polynomiality.

Polynomiality follows from Poincar\'e symmetry
(and time reversal symmetry), but not vice versa: 
polynomiality is a necessary but not sufficient condition 
for the compliance of a model with Poincar\'e invariance
\cite{Ji:1997gm}.
A consistent description of the energy-momentum tensor form 
factors $A(t)$ and $J(t)$ \cite{Polyakov:2018zvc}
in the bag model can be obtained if and only if one sets
$\eta=\eta_0$ as defined in the sequence of 
Eq.~(\ref{Eq:tildeDelta}) \cite{Ji:1997gm}.
Due to Poincar\'e invariance, these form factors must satisfy 
at $t=0$ the constraints $A(0)=1$ and $J(0)=\frac12$,
see e.g.\ \cite{Cotogno:2019xcl}.
The first constraint holds in the bag model for any $\eta$,
but the second requires $\eta=\eta_0$. 
Technically, this is because $A(t)$ appears without and 
$J(t)$ with a prefactor of $\Delta^\mu$ in the Lorentz
decomposition of the energy-momentum tensor matrix elements.
While one is free to rescale the momentum transfer which 
goes into the ``internal'' evaluation in a model 
in Eq.~(\ref{Eq:tildeDelta}), this is of course not 
possible for an ``external'' factor $\Delta^\mu$ from 
the Lorentz decomposition. 
As a result of this mismatch, form factors without $\Delta^\mu$
in the Lorentz decomposition are in general $\eta$-independent
at $t=0$, and those with exhibit a residual $\eta$-dependence.
(This is true also for the tensor form factors: $H_T(0)$ does
not, while $E_T(0)$ and $\tilde{H}_T(0)$ do depend on $\eta$.
None of these form factors is however constrained by Poincar\'e
symmetry.)
The failure to comply with $J(0)=\frac12$ for $\eta\neq\eta_0$
signals non-compliance with Poincar\'e invariance 
\cite{Ji:1997gm} --- even though polynomiality remains
valid for any value of $\eta$.

Let's keep in mind that the step to allow $\eta\neq\eta_0$ 
was introduced to remedy, in an independent particle model, the
calculation of form factors at $t\neq 0$ \cite{Ji:1997gm}, 
see the end of Sec.~\ref{Sec-3:bag-model}.
This is of course not adequate for a correct 
description of model matrix elements for $t\to0$.
A more consistent modeling could perhaps be achieved 
by working with a function $\eta=\eta(t)$ chosen such 
that $\eta(0)=\eta_0$. 
Based on an appropriate form for $\eta(t)$ the bag model 
could then comply with the constraint $J(0)=\frac12$, polynomiality,
and a good phenomenological description of electromagnetic
form factors.
It would be interesting to investigate such an approach
in future work. While presumably more consistent, such
a procedure would remain a rough method to remedy the
shortcomings of independent particle models.

A different way to circumvent the shortcomings of independent particle
models is to work in the large-$N_c$ limit \cite{Neubelt:2019sou}.

\end{document}